\documentclass[aps,twocolumn,superscriptaddress,floatfix,longbibliography]{revtex4-2}
\usepackage{amsmath,amssymb,amsthm}
\usepackage{physics}
\usepackage{amsfonts}
\usepackage{mathrsfs}
\usepackage{graphicx}
\usepackage{tabularx}
\usepackage{enumerate}
\usepackage{dcolumn}
\usepackage{bm}
\usepackage{xcolor}
\usepackage[normalem]{ulem}
\usepackage[colorlinks,linkcolor=blue,citecolor=blue,urlcolor=blue]{hyperref}

\begin{document}
	\preprint{APS/123-QED}

\title{Quantum Pontus--Mpemba Effect in Dissipative Quasiperiodic Chains}

\author{Yefeng Song}
\affiliation{School of Physics, Nankai University, Tianjin 300071, China}

\author{Junxiao Chen}
\affiliation{School of Physics, Nankai University, Tianjin 300071, China}

\author{Xiangyu Yang}
\affiliation{School of Physics, Nankai University, Tianjin 300071, China}

\author{Mingdi Xu}
\affiliation{School of Physics, Nankai University, Tianjin 300071, China}
	\author{Xiang-Ping Jiang}
\email{2015iopjxp@gmail.com}
\affiliation{School of Physics, Hangzhou Normal University, Hangzhou, Zhejiang 311121, China}

\author{Lei Pan}
\email{panlei@nankai.edu.cn}
\affiliation{School of Physics, Nankai University, Tianjin 300071, China}


\date{\today}

\begin{abstract}
We investigate how quasiperiodic spatial structure enables protocol-induced acceleration in open quantum systems by analyzing the Pontus–Mpemba effect in one-dimensional chains subject to Markovian dephasing. The dynamics are governed by a Lindblad superoperator that drives all initial states toward a maximally mixed infinite-temperature steady state, isolating dynamical mechanisms from static equilibrium properties.
Considering two representative quasiperiodic models, namely a tight-binding chain with a mosaic potential and its extension with power-law long-range hopping, we show that a properly engineered two-step protocol, in which the system is first steered to a finite-temperature intermediate state, yields a strictly shorter overall relaxation time than direct evolution from the same initial configuration.
This protocol-induced acceleration persists for both initially localized and extended eigenstates and remains robust in the presence of long-range hopping. A Liouvillian spectral analysis reveals that the mechanism originates from a redistribution of spectral weight that suppresses overlap with the slowest decay modes, rather than from any modification of the decay spectrum itself.
Our results establish quasiperiodic chains as a controlled setting for engineering relaxation pathways through Liouvillian spectral structure.
\end{abstract}

\maketitle

\section{Introduction}

The relaxation of physical systems toward equilibrium is a fundamental problem in statistical mechanics. Under conventional intuition, the relaxation time is expected to decrease monotonically as the initial state approaches equilibrium. The Mpemba effect defies this expectation: when two systems are quenched under identical cooling conditions, the initially hotter one may relax to the colder equilibrium state in a shorter time than the initially cooler one. First reported by Mpemba and Osborne~\cite{ME}, this counterintuitive phenomenon has since been observed in diverse classical settings, including granular media~\cite{ME_classical9}, colloidal suspensions~\cite{ME_classical5}, molecular systems~\cite{ME_classical7,ME_classical8}, and related experimental platforms~\cite{ME_classical1,ME_classical2,ME_classical3,ME_classical4,ME_classical6}. An inverse variant, in which colder systems heat faster than warmer ones,
has also been reported~\cite{Inverse_ME1,Inverse_ME2,Inverse_ME3}.

The exploration of analogous behavior in the quantum regime
has given rise to what is now termed the quantum Mpemba effect (QME). Experimental realizations in controllable quantum platforms~\cite{QME_Exp1,QME_Exp2,QME_Exp3,QME_Exp4} have stimulated extensive theoretical investigations in integrable systems~\cite{QME1,QME101,QME2,QME3,QME4,QME41}, disordered and many-body localized phases~\cite{QME5}, random quantum circuits~\cite{QME8,QME9,QME901,QME902,QME903,QME904}, quantum dots~\cite{QME_dot1,QME_dot2}, and open quantum systems described by Lindblad master equations~\cite{OpenQME1,OpenQME2,OpenQME3,OpenQME4,OpenQME5,OpenQME6,OpenQME7,OpenQME8,OpenQME10,OpenQME11,OpenQME12,OpenQME13,wei2025quantum,caldas2025exponentially,liu2025general}. Related studies have further explored non-Hermitian systems~\cite{QME10,QME105}, quantum harmonic oscillators~\cite{QME12}, Sachdev–Ye–Kitaev models~\cite{QME15}, and control-based or resource-theoretic formulations~\cite{QME20,QME21,QME26,QME30,QME31}.

In open quantum systems, relaxation toward the steady state is governed by the spectral structure of the Liouvillian superoperator~\cite{R1,R2}. The asymptotic dynamics are controlled by eigenmodes with the smallest nonzero real parts, commonly characterized by the Liouvillian gap. Within this spectral framework, the QME can be interpreted as arising from the nontrivial dependence of relaxation on the projection of the initial state onto slow Liouvillian decay modes~\cite{OpenQME1,QME_Exp3,OpenQME7,wei2025quantum}. The rapid progress in engineering dissipative quantum platforms~\cite{Exp1,Exp2,Exp3,Exp4,Exp5,Exp6,Exp7,Exp8,Exp_new1,Exp_new2,Exp_new3} has therefore intensified interest in identifying and controlling anomalous relaxation mechanisms in many-body open systems.

A recent conceptual extension, termed the Pontus-Mpemba effect (PME)~\cite{QME27}, shifts the emphasis from comparing distinct initial states to designing alternative dynamical protocols. Rather than fixing the evolution generator and varying the initial condition, PME contrasts different dynamical routes that originate from the same initial state. In a typical two-step protocol, the system first evolves under an auxiliary Hamiltonian or dissipative environment to produce an intermediate nonequilibrium state, after which the target dynamics are restored. A PME is identified when the total relaxation time under this composite evolution is shorter than that of direct evolution toward the same steady state. This strategy has been demonstrated in Markovian open systems~\cite{QME27,Longhi_2026}, in systems exhibiting dissipative phase transitions~\cite{QME28}, and even in closed systems under real- and imaginary-time dynamics~\cite{QME29}, establishing intermediate-state engineering as a systematic route to accelerated quantum relaxation. While the QME is typically analyzed in terms of initial-state projections onto slow Liouvillian modes, PME can be viewed as a protocol that actively reshapes these projections through an intermediate evolution stage.

At the same time, dissipative dynamics in non-ergodic quantum systems have attracted considerable attention, including systems with random disorder~\cite{Yusipov17,Yusipov18,Jiang_3D}, incommensurate potentials~\cite{WYC_PRL,Xu_FlatBand,feng2025localization,roy2025aperiodic}, many-body localization~\cite{WYC_MBL}, and quantum scar states~\cite{Diss_Scar1,Diss_Scar2,Diss_scar3,Diss_scar4,Diss_scar5}. Recent works suggest that anomalous heating and cooling behaviors may persist in such settings~\cite{QME6,QME7,QME705}, indicating that quasiperiodicity and dissipation can jointly reshape relaxation pathways. In particular, accelerated thermalization in incommensurate systems~\cite{QME7} points to a nontrivial interplay between structural inhomogeneity and Liouvillian spectral properties.

In this work, we analyze the emergence of the PME in one-dimensional quasiperiodic lattices subject to Markovian dephasing. We consider two representative models: a nearest-neighbor tight-binding chain with a quasiperiodic mosaic potential~\cite{Mosaic1}, and its extension with power-law long-range hopping under analogous quasiperiodic modulation~\cite{LongRange1}. The dissipative dynamics are described by a Lindblad master equation~\cite{Lindblad1,Lindblad2}, where local dephasing drives the system toward the maximally mixed steady state.
We demonstrate that, in both models and across regimes featuring localized and extended eigenstates, a suitably designed two-step Pontus protocol yields a strictly shorter overall relaxation time than direct evolution. By resolving the Liouvillian eigenmode decomposition, we show that the acceleration originates from a controlled redistribution of spectral weight away from slow Liouvillian modes during the intermediate stage, rather than from any modification of the decay spectrum itself. Our results establish that Pontus-type acceleration persists in quasiperiodic systems with mobility edges and long-range hopping, and provide a unified Liouvillian perspective on anomalous relaxation in structurally inhomogeneous open quantum systems.

	\section{THEORETICAL FRAMEWORK }\label{sec: models}
A convenient starting point for analyzing Mpemba-type phenomena in quantum systems
is the dissipative dynamics of open systems.
We consider a system coupled to an environment,
described by the total Hamiltonian
\begin{equation}
	H = H_S + H_B + H_{SB},
\end{equation}
where $H_S$ and $H_B$ denote the system and bath Hamiltonians,
and $H_{SB}$ their interaction.
Under the Born–Markov approximation~\cite{Moy1999,Breuer2002},
and after tracing out the bath degrees of freedom,
the reduced system dynamics is governed by the Lindblad master equation~\cite{Lindblad1,Lindblad2},
\begin{equation}
	\frac{d\rho(t)}{dt}
	=
	\mathscr{L}[\rho(t)]
	=
	-i[H_S,\rho(t)] + \mathcal{D}[\rho(t)],
	\label{eq:Lindblad}
\end{equation}
where $\mathscr{L}$ is the Liouvillian superoperator generating completely positive and trace-preserving dynamics. The dissipator takes the standard form
\begin{equation}
	\mathcal{D}[\rho]
	=	\sum_{i}\sum_{m=1}^{M}
	\Gamma_i^{(m)}
	\Big(	O_i^{(m)} \rho O_i^{(m)\dagger}	-\frac{1}{2}\left\{ O_i^{(m)\dagger} O_i^{(m)}, \rho \right\}\Big),
\end{equation}
with jump operators $O_i^{(m)}$ acting on site $i$ and rates $\Gamma_i^{(m)}$.

The dissipative evolution is fully characterized by the spectral properties 
of the Liouvillian superoperator. The time-dependent density matrix can be written as
\begin{equation}
	\rho(t)=e^{\mathscr{L}t}\rho(0).
\end{equation}
In the long-time limit, the system relaxes to the steady state 
$\rho_{\mathrm{ss}}$, which corresponds to the right eigenoperator of 
$\mathscr{L}$ with eigenvalue $\lambda_1=0$. Expanding in right and left eigenmodes, one obtains
\begin{eqnarray}
	\rho(t)
	=
	\rho_{\mathrm{ss}}
	+
	\sum_{n=2}^{D^2}
	\mathrm{Tr}\left[l_n\rho(0)\right]
	r_n
	e^{\lambda_n t},
	\label{eq:timeevolution}
\end{eqnarray}
where $\mathscr{L} r_n=\lambda_n r_n$ and $\mathscr{L}^\dagger l_n=\lambda_n^* l_n$. All nonzero eigenvalues satisfy $\mathrm{Re}(\lambda_n)<0$. The eigenvalue with the smallest magnitude real part, $\lambda_2$, defines the Liouvillian gap and controls the asymptotic relaxation timescale.

Equation~\eqref{eq:timeevolution} makes explicit that relaxation is governed not only by the spectral gap but also by the overlap $\mathrm{Tr}[l_n\rho(0)]$ between the initial state and the decay modes. Suppressing the projection onto the slowest mode $r_2$ can therefore lead to faster effective relaxation--providing the spectral basis for Mpemba-type effects.
In the relaxation stage we consider local dephasing with jump operators $O_i^{(m)}=n_i$ with dissipation strength $\Gamma_i^{(m)}=\Gamma_1$, where $n_i$ is the on-site density operator. This dissipative channel drives the system toward the infinite-temperature state
\begin{equation}
	\rho_{\mathrm{ss}}=\frac{1}{D}\mathbf{I},
\end{equation}
with $D$ the Hilbert-space dimension.

To implement the first stage of the Pontus protocol, we transiently couple the system to a finite-temperature bosonic environment, which drives it toward a thermal state of the system Hamiltonian. 
Suppose the interaction Hamiltonian is written as $H_{SB}=S\otimes B$, in the eigenbasis of
\begin{equation}
	H_S=\sum_n E_n |n\rangle\langle n|,
\end{equation}
the system operator $S$ is decomposed into components associated with frequencies,
\begin{equation}
	S=\sum_{\omega} A_\omega,
\end{equation}
where
\begin{equation}
	A_\omega=\sum_{E_m-E_n=\omega}
	|n\rangle\langle n|\, S \,|m\rangle\langle m|.
\end{equation}

Under the Born-Markov approximations, the reduced dynamics is governed by a thermal Lindblad dissipator of the form~\cite{Breuer2002}
\begin{equation}
	\mathcal{D}_{\mathrm{th}}[\rho]
	=
	\sum_{\omega}
	\gamma(\omega)
	\left(
	A_\omega \rho A_\omega^\dagger
	-
	\frac{1}{2}
	\left\{A_\omega^\dagger A_\omega,\rho\right\}
	\right).
\end{equation}

We consider an Ohmic bosonic bath characterized by the spectral density
\begin{equation}
	J(\omega)
	=
	\eta\,\omega\, e^{-\omega/\omega_c},
	\quad \omega>0,
\end{equation}
which is a standard model for finite-temperature dissipation in quantum optical and condensed-matter settings~\cite{Breuer2002}. 
The corresponding transition rates are
\begin{equation}
	\gamma(\omega)
	=
	J(|\omega|)
	\begin{cases}
		n_B(|\omega|)+1, & \omega>0, \\
		n_B(|\omega|), & \omega<0,
	\end{cases}
\end{equation}
where $n_B(\omega)=1/(e^{\beta\omega}-1)$ is the Bose distribution. 
These rates satisfy the detailed-balance condition
\begin{equation}
	\frac{\gamma(+\omega)}{\gamma(-\omega)}=e^{\beta\omega},
\end{equation}
which guarantees a unique Gibbs steady state
\begin{equation}
	\rho_{\mathrm{th}}
	=
	\frac{e^{-\beta H_S}}
	{\mathrm{Tr}\left(e^{-\beta H_S}\right)} .
\end{equation}

Within the two-stage Pontus protocol, this thermal bath acts only during a finite preparation interval, steering the system toward $\rho_{\mathrm{th}}$ at temperature $T=1/\beta$. After this intermediate state is generated, the bath is removed and the subsequent evolution proceeds under pure dephasing toward the infinite-temperature steady state. Since the Liouvillian governing the second stage is identical for all preparations, any difference in relaxation speed arises solely from the distinct projections of the prepared states onto the decay eigenmodes of the dephasing Liouvillian. This construction therefore provides a fully Markovian and spectrally controlled setting for isolating the mechanism underlying the PME.

\section{Numerical results}\label{sec: Numerical}

To quantify relaxation toward the steady state $\rho_{\mathrm{ss}}$, 
we employ the trace distance
\begin{equation}
	D_{\mathrm{tr}}(\rho,\rho_{\mathrm{ss}})
	=
	\frac{1}{2}\,\mathrm{Tr}
	\sqrt{(\rho-\rho_{\mathrm{ss}})^\dagger(\rho-\rho_{\mathrm{ss}})} .
\end{equation}
The trace distance is monotonic under completely positive trace-preserving maps and vanishes if and only if $\rho=\rho_{\mathrm{ss}}$, 
making it a natural metric for dissipative convergence.
In the asymptotic regime, its decay is controlled by the Liouvillian 
eigenmodes with the smallest nonzero real parts, which dominate the late-time dynamics.

\subsection{mosaic model}\label{sec: mosaic}

\begin{figure}[t]
	\centering
	\includegraphics[width=\linewidth]{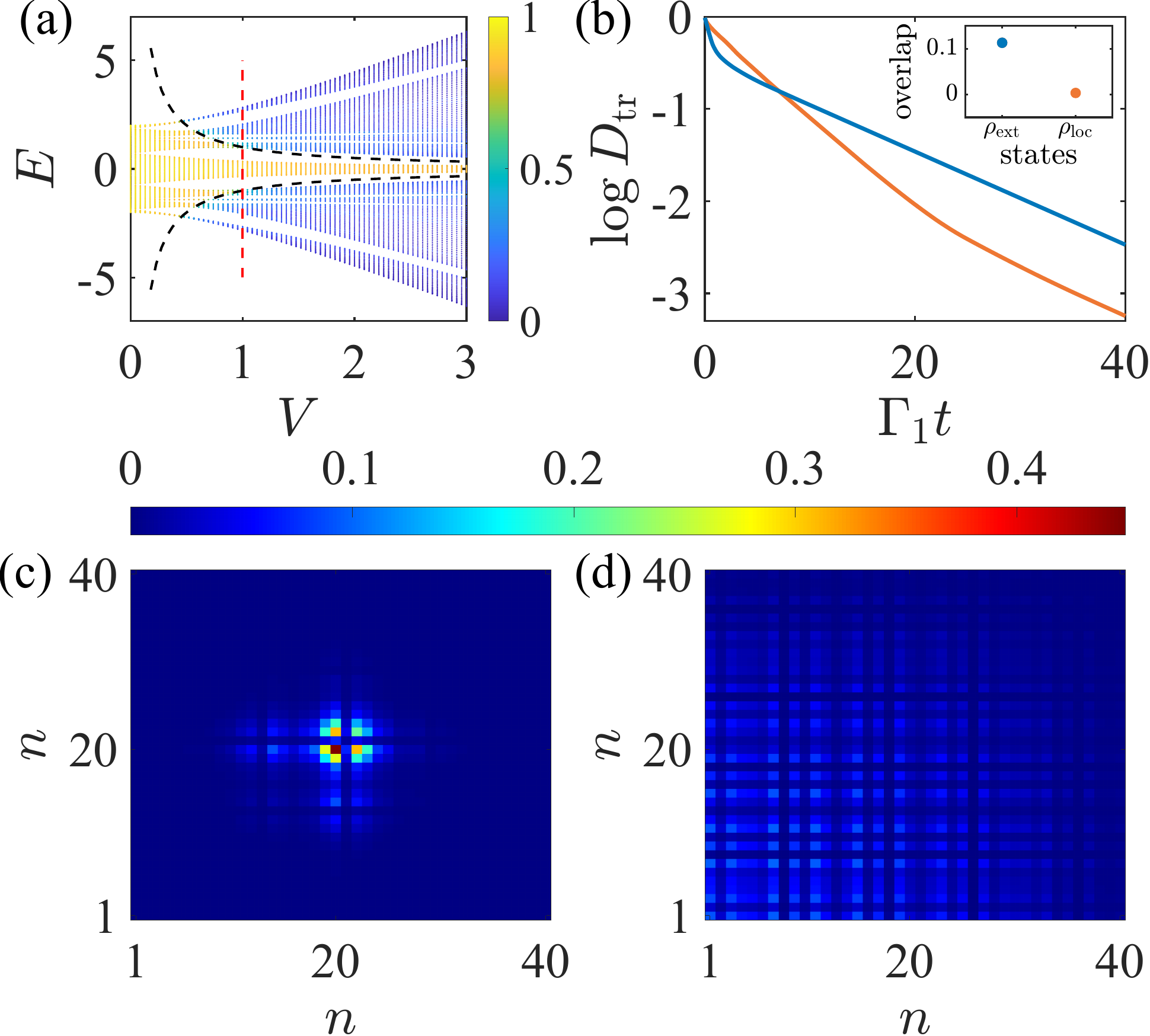}
	\caption{
		(a) Fractal dimension $D_q$ as a function of eigenenergy $E$ and quasidisorder strength $V$.
		Black dashed lines indicate the analytical mobility edges,
		and the red dashed line marks the critical point $V=1$.
		(b) Time evolution of the trace distance $D_{\mathrm{tr}}$ under pure dephasing
		for a localized eigenstate ($E=-2.03$, orange) and an extended eigenstate ($E=0.35$, blue).
		Inset: overlap of each eigenstate with the slowest Liouvillian decay mode.
		(c),(d) Spatial density distributions corresponding to the localized and extended eigenstates in (b).
		Parameters: $\beta=(\sqrt{5}-1)/2$, $t=1$, $V=1$, $\Gamma_1=0.1$.
		System size: $L=610$ in (a) and $L=40$ in (b)–(d).
	}
	\label{fig1}
\end{figure}

We begin with the one-dimensional mosaic model~\cite{Mosaic1},
a quasiperiodic lattice described by
\begin{equation}
	H = -t \sum_i \left(c_i^\dagger c_{i+1} + c_{i+1}^\dagger c_i\right)
	+ 2 \sum_i V_i c_i^\dagger c_i ,
\end{equation}
where $t$ denotes the nearest-neighbor hopping amplitude.
The on-site potential takes the form
$V_i = V \cos(2\pi \beta i + \phi)$ on even sites and vanishes on odd sites,
with irrational $\beta$ and phase $\phi$.
This model possesses exact mobility edges at
$E_{\mathrm{ME}}=\pm t/V$,
separating extended states in the central region of the spectrum
from localized states in the spectral wings.

Localization properties are quantified through the inverse participation ratio
$\mathrm{IPR}(k)=\sum_i |\psi_{k,i}|^4$,
from which the fractal dimension
$D_q = -\lim_{L\to\infty}\ln(\mathrm{IPR})/\ln L$
is extracted.
Extended states yield $D_q\approx1$,
while localized states give $D_q\approx0$.
Figure~\ref{fig1}(a) confirms the sharp transition across the mobility edges.
At the critical value $V=1$, states with $|E|>1$ are localized,
whereas those with $|E|<1$ remain extended.

We now turn to dissipative dynamics under local dephasing with rate $\Gamma_1=0.1$.
Figure~\ref{fig1}(b) compares the relaxation of the 6th eigenstate
($E=-2.03$, localized)
and the 25th eigenstate
($E=0.35$, extended)
for a system of size $L=40$.
Both states are prepared with identical initial trace distance
from the steady state $\rho_{\mathrm{ss}}$.
Despite being spatially more concentrated
[Figs.~\ref{fig1}(c),(d)],
the localized state relaxes more rapidly at intermediate and long times.
This behavior does not follow from spatial structure alone,
but is rooted in the Liouvillian spectrum.
As shown in the inset of Fig.~\ref{fig1}(b),
the localized eigenstate has a smaller overlap with the slowest decay mode.
Since the long-time dynamics is dominated by this mode,
the reduced projection leads to a faster effective convergence to the steady state.

\begin{figure}[t]
	\centering
	\includegraphics[width=\linewidth]{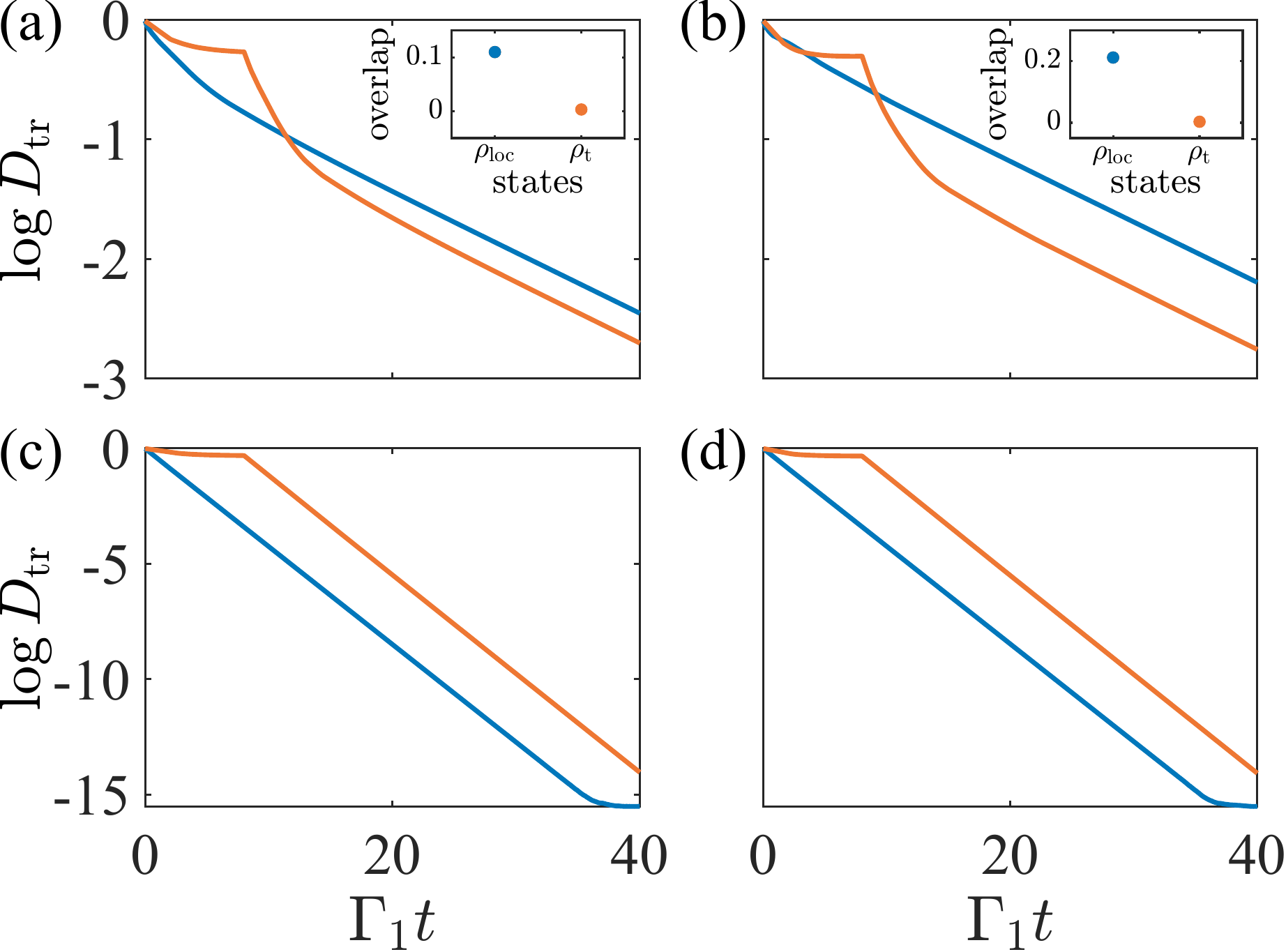}
\caption{
	PME in the localized regime.
	(a), (b) Time evolution of the trace distance $D_{\mathrm{tr}}$
	for the two-step protocol (orange: weak coupling to a finite-temperature Ohmic bath)
	and the direct one-step protocol (blue: pure dephasing).
	The prethermalization duration is chosen as one-fifth of the total evolution time, with a weak auxiliary coupling $\Gamma_2=\Gamma_1/100$ and bath temperature $T=1$.
	Panel (a): $7\mathrm{th}$ eigenstate ($E=-1.98$, $T_{\mathrm{eff}}=1.10$);
	panel (b): $12\mathrm{th}$ eigenstate ($E=-1.18$, $T_{\mathrm{eff}}=2.35$).
	Insets show the overlaps of the initial and intermediate states
	with the slowest Liouvillian decay mode.
	(c), (d) Corresponding control results with the quasiperiodic potential removed ($V=0$).
	Other parameters are identical to Fig.~\ref{fig1}.
}
	\label{fig2}
\end{figure}

Motivated by the mode-structure analysis above,
we implement a two-step control protocol.
During an initial interval $0<t<\tau_1$,
the system is weakly coupled to a finite-temperature bath
with rate $\Gamma_2=\Gamma_1/100$,
allowing controlled prethermalization.
At $t=\tau_1$ the auxiliary bath is switched off,
and pure dephasing with rate $\Gamma_1$ governs the subsequent evolution.
For comparison, the one-step protocol applies dephasing throughout.

We first consider eigenstates in the localized regime ($|E|>1$).
Figure~\ref{fig2}(a) shows the dynamics of the 7th eigenstate
($E=-1.98$, $T_{\mathrm{eff}}=1.10$),
with the auxiliary bath temperature chosen as $T=1<T_{\mathrm{eff}}$. For each eigenstate $|\psi_n\rangle$ with energy $E_n$, we define an effective temperature $T_{\mathrm{eff}}$ by equating its energy expectation value to that of a canonical ensemble, $E_n = \frac{\mathrm{Tr}\left( H_S e^{-H_S/T_{\mathrm{eff}}} \right)}{\mathrm{Tr}\left( e^{-H_S/T_{\mathrm{eff}}} \right)}$.
This definition provides a convenient thermodynamic interpretation of the eigenstate energy, allowing us to compare it with the temperature of the auxiliary bath used in the first stage of the protocol.
Although the two-step evolution initially relaxes more slowly during the prethermalization stage,
a clear crossover occurs after the bath is removed,
beyond which the convergence becomes systematically faster than in the direct dephasing case.

A similar behavior is observed for other localized eigenstate
under the same protocol [Fig.~\ref{fig2}(b)]. In both examples,
the intermediate thermal state generated during prethermalization
exhibits a reduced overlap with the slowest Liouvillian mode,
thereby suppressing the dominant long-time contribution.

To assess the role of quasidisorder, we repeat the same procedure after removing the on-site potential ($V=0$). As shown in Figs.~\ref{fig2}(c) and (d), the crossover disappears in the clean limit, and the one-step evolution remains faster throughout. This comparison demonstrates that the acceleration relies on the nontrivial Liouvillian structure induced by the quasiperiodic potential.


\begin{figure}[t]
	\centering
	\includegraphics[width=\linewidth]{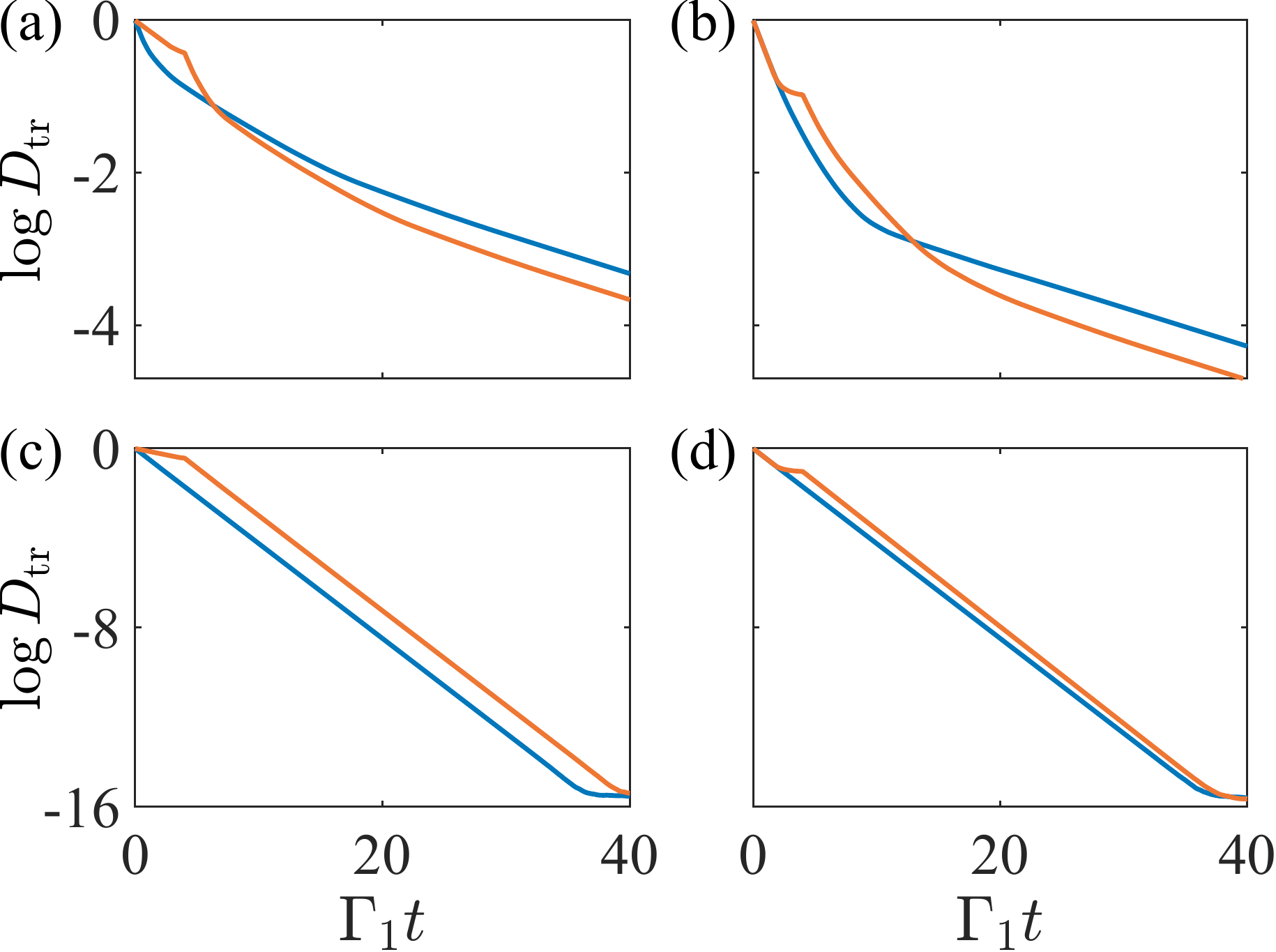}
	\caption{
		PME in the extended regime.
		(a) Dynamics of the 17th eigenstate ($E=-0.33$, $T_{\mathrm{eff}}=10.43$)
		prethermalized by a bath at $T=2$.
		(b) Dynamics of the 20th eigenstate ($E=-0.08$, $T_{\mathrm{eff}}=135.74$)
		prethermalized at $T=7$.
		In both cases the prethermalization duration is chosen as one-tenth of the total evolution time, 
		with a weak auxiliary coupling $\Gamma_2=\Gamma_1/250$ and bath temperature $T=1$.	(c),(d) Corresponding results in the clean limit ($V=0$).
		Orange (blue) curves denote the two-step (one-step) scheme.
	}
	\label{fig3}
\end{figure}

We next examine eigenstates in the extended regime ($|E|<1$),
where wave functions are spatially delocalized and transport is ballistic at the Hamiltonian level. It is therefore not a priori clear whether the acceleration observed above persists in this regime.

Figure~\ref{fig3}(a) presents the evolution of extended eigenstate
with energy $E=-0.33$ and effective temperature $T_{\mathrm{eff}}=10.43$. The auxiliary bath temperature is chosen as $T=2$, significantly lower than $T_{\mathrm{eff}}$.
The prethermalization strength is reduced to $\Gamma_2=\Gamma_1/250$. Despite these weaker parameters,
a clear crossover in the trace distance again develops: after the prethermalization stage, the two-step trajectory converges faster than direct dephasing.

An analogous behavior is observed for a highly excited eigenstate,
the 20th state with $E=-0.08$ and $T_{\mathrm{eff}}=135.74$,
when prethermalized at $T=7$ [Fig.~\ref{fig3}(b)]. The acceleration thus persists over a wide energy window, including states with extremely high effective temperatures.
In contrast, when the quasiperiodic potential is removed ($V=0$),
the crossover disappears in both cases [Figs.~\ref{fig3}(c),(d)].
The one-step protocol remains uniformly faster. This comparison confirms that the PME does not arise from trivial temperature reduction alone, but from the interplay between quasidisorder and the structure of Liouvillian decay modes.

Taken together, the results obtained for both localized and extended eigenstates demonstrate that the acceleration under the two-step protocol is not confined to a specific spectral sector, but persists across distinct dynamical regimes. 
Since the Liouvillian spectrum remains unchanged throughout the protocol, the observed acceleration cannot originate from a modification of the decay rates themselves. Instead, it must result from a redistribution of spectral weight among Liouvillian eigenmodes.


\begin{figure*}[t]
	\centering
	\includegraphics[width=\linewidth]{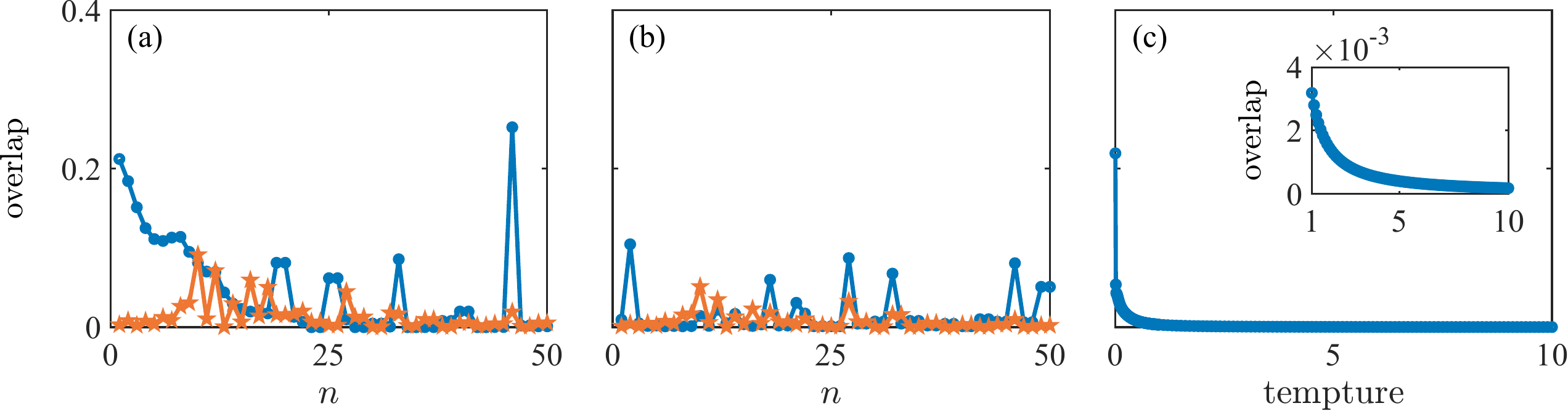}
	\caption{
		Spectral mechanism underlying the acceleration.
		(a),(b) Absolute overlaps between Liouvillian decay modes
		and two states:
		the initial eigenstate (blue)
		and the intermediate thermal state (orange).
		Panel (a) corresponds to the 12th eigenstate
		and a thermal state at $T=1$;
		panel (b) to the 17th eigenstate
		and a thermal state at $T=2$.
		(c) Overlap between thermal states at different temperatures
		and the slowest decay mode.
		Inset: $1\le T\le10$.
	}
	\label{fig4}
\end{figure*}

To elucidate the microscopic origin of the acceleration, we analyze the spectral decomposition of the Liouvillian superoperator. Let $\{r_n\}$ and $\{l_n\}$ denote its right and left eigenmodes, ordered according to increasing decay rate, with eigenvalues $\lambda_n$ satisfying $\mathrm{Re}(\lambda_1)=0 > \mathrm{Re}(\lambda_2) \ge \mathrm{Re}(\lambda_3)\dots\ge \mathrm{Re}(\lambda_{D^2})$. In the long-time regime, the dynamics is dominated by the slowest nonzero mode $r_2$, whose contribution is weighted by the overlap $|\mathrm{Tr}(l_2 \rho_0)|$ with the initial state. The key effect of the prethermalization stage is to reshape this projection: the intermediate thermal state exhibits a significantly reduced overlap with $l_2$, thereby suppressing the slowest decay channel and accelerating the subsequent relaxation toward the steady state.

Figures~\ref{fig4}(a) and (b) compare these overlaps for representative localized and extended states. In panel (a), we consider the 12th eigenstate ($E=-1.18$) and the corresponding thermal state at $T=1$. In panel (b), we analyze the 17th eigenstate ($E=-0.33$) and a thermal state at $T=2$. In both cases, the thermal states generated during prethermalization exhibit substantially smaller projections onto the slowest decay modes than the original eigenstates.

Since the asymptotic relaxation rate is fixed by the Liouvillian gap,
the only way to accelerate convergence is to suppress the initial weight carried by these slow modes. The two-step protocol accomplishes precisely this: it redistributes spectral weight away from dynamical bottlenecks, thereby shortening the effective relaxation time.

Figure~\ref{fig4}(c) further shows the temperature dependence of the overlap between thermal states and the slowest decay mode.
For $1\le T\le10$ (inset), the overlap decreases monotonically with increasing temperature, and remains small over a broad range.
This behavior explains why intermediate thermal states prepared at moderate temperatures can systematically accelerate relaxation
even when their initial trace distance from the steady state is larger.

The PME observed in this model therefore originates from a controlled redistribution of spectral weight among Liouvillian eigenmodes, in particular a reduction of the projection onto
the slowest decay mode, rather than from any modification of the decay spectrum itself. The quasiperiodic potential plays a central role
by generating a structured Liouvillian spectrum in which slow modes are well separated and spectrally identifiable.

\subsection{Power-law hopping model}
\label{sec: Power-law hopping}

We next consider a generalized Aubry--Andr\'{e} (GAA) model with long-range power-law hopping~\cite{LongRange1}, described by
\begin{equation}
	H = -t\sum_{i,j\neq i}\frac{1}{|i-j|^a}\,c^\dagger_i c_j 
	+ V\sum_i\cos\!\left[\beta(2\pi i+\phi)\right] c^\dagger_i c_i ,
\end{equation}
where $c_i^\dagger$ ($c_i$) creates (annihilates) a particle at site $i$, $t$ denotes the hopping amplitude, and $\beta=(\sqrt{5}-1)/2$.
In the nearest-neighbor limit ($a\to\infty$), the model reduces to the standard Aubry--Andr\'{e} Hamiltonian, which is self-dual and exhibits a localization transition at $V=2t$ separating an all-ergodic (AE) phase from an all-localized (AL) phase. 

\begin{figure}[t!]
	\centering
	\includegraphics[width=1\linewidth]{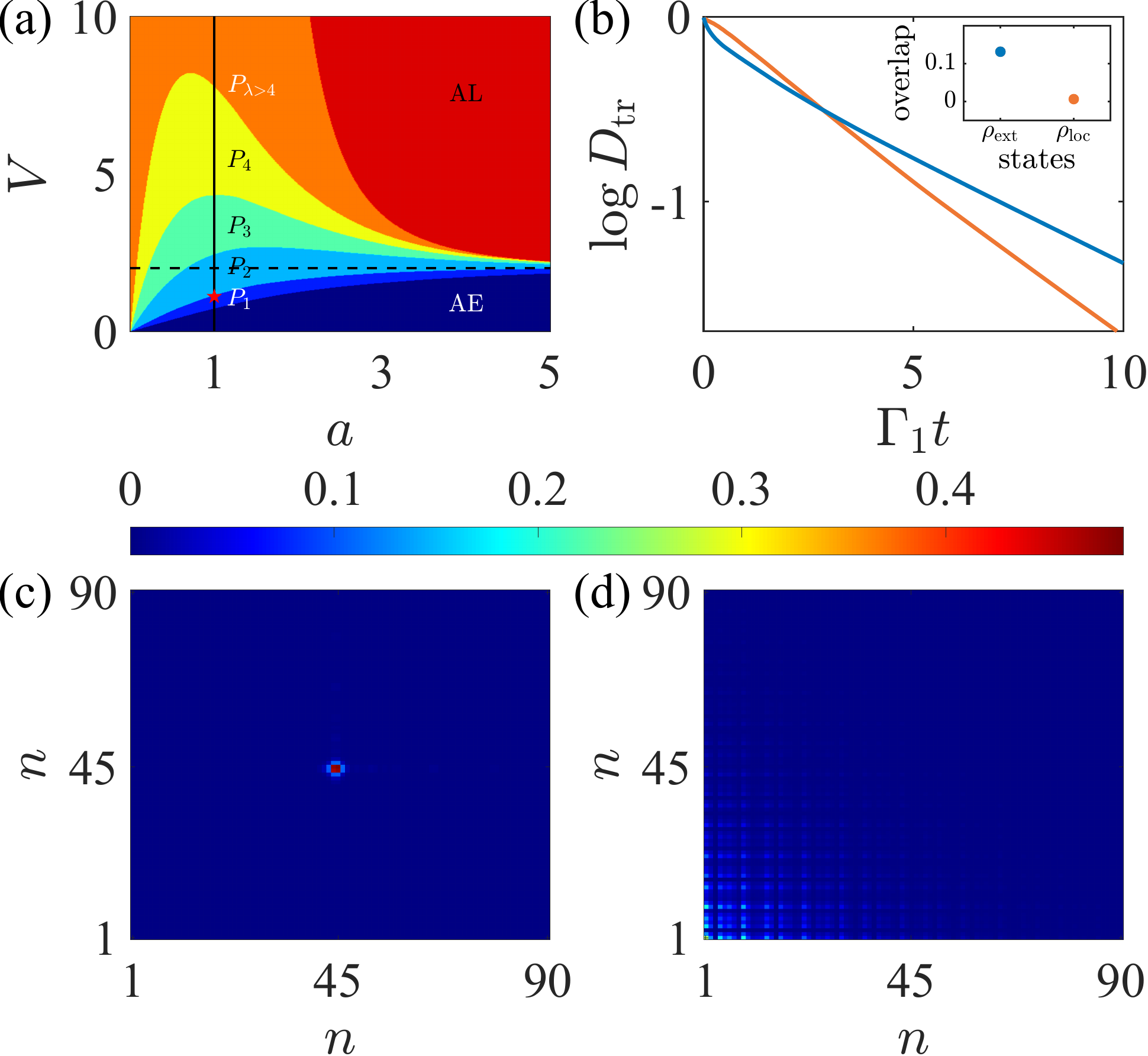}
	\caption{
		Relaxation dynamics in the generalized Aubry--Andr\'{e} model with power-law hopping.
		(a) Phase diagram in the $(V,a)$ plane for $L=987$ and $t=1$.
		The dashed line denotes the conventional AA boundary $V=2t$,
		while the solid horizontal line marks $a=1$.
		The red star indicates the parameter set used in panels (b)--(d).
		(b) Time evolution of the trace distance $D_{\mathrm{tr}}$ under pure dephasing
		with rate $\Gamma_1=0.1$,
		starting from a localized eigenstate
		(22\text{nd}, orange)
		and an extended eigenstate
		(57\text{nd}, blue)
		at $L=90$, $t=1$, and $V=1.1$.
		A crossover in the relaxation curves is observed,
		signaling Mpemba-like behavior.
		The inset shows the projection of each initial state onto the slowest Liouvillian decay mode.
		(c),(d) Spatial particle-density profiles
		of the localized and extended eigenstates in (b), respectively.
		Here $\beta=(\sqrt{5}-1)/2$ and periodic boundary conditions are used.
	}
	\label{fig5}
\end{figure}

Introducing power-law hopping ($a<\infty$) breaks self-duality and generates a considerably richer phase diagram.
Figure~\ref{fig5}(a) presents the phase structure in the $(V,a)$ plane for $L=987$ and $t=1$. The dashed line marks the conventional AA boundary $V=2t$, while the solid horizontal line indicates $a=1$.
For $a>1$, the hopping is effectively short-ranged and the phase diagram develops a staircase structure. Between the AE phase at small $V$ and the AL phase at large $V$, a hierarchy of intermediate phases $P_s$ emerges.
Within each $P_s$ phase, mobility edges partition the spectrum: the lowest $\sim \beta^s L$ eigenstates remain ergodic,
whereas higher-energy states are localized. Consequently, increasing $V$ induces a stepwise localization process in which the fraction of localized states grows discontinuously. In the limit $a\gg1$, long-range hopping is suppressed and the model recovers the AA universality class with a sharp transition at $V=2t$.

For $a\le1$, however, long-range hopping qualitatively alters the localization properties. The fully localized phase disappears, and the high-energy states in the $\mathrm{P}_s$ phases become multifractal rather than exponentially localized, characterized by a fractal dimension $0<D_q<1$. In this regime, the phase boundary separates ergodic and multifractal states instead of ergodic and localized ones. In the following we focus on the representative point
$L=90$, $t=1$, $V=1.1$, and $a=1$ (indicated by the red star in Fig.~\ref{fig5}(a)), which lies inside the $P_1$ phase.

We now examine the dissipative dynamics at the representative parameter point $L=90$, $t=1$, $V=1.1$, and $a=1$, located in the $P_1$ phase of the phase diagram. Figure~\ref{fig5}(b) shows the time evolution of the trace distance $D_{\mathrm{tr}}$ from the infinite-temperature steady state $\rho_{\mathrm{ss}}$
under pure dephasing with rate $\Gamma_1=0.1$. We compare two eigenstates that coexist in the same spectrum: the 22\text{nd} eigenstate (localized or strongly multifractal, orange)
and the 57\text{th} eigenstate (extended, blue). 
Although the two states are prepared with identical initial trace distance, their subsequent relaxation dynamics differ substantially.
Following a short transient regime, the localized state exhibits a faster decay at intermediate and long times, eventually overtaking the extended state. This inversion of relaxation ordering demonstrates a Mpemba-like effect in a quasiperiodic system featuring mobility edges and long-range hopping. The result highlights that the relaxation rate is not determined solely by the initial distance to the steady state, but rather by the spectral decomposition of the initial state onto slow Liouvillian modes.

The real-space density profiles shown in Figs.~\ref{fig5}(c) and \ref{fig5}(d) reveal that the localized eigenstate is strongly inhomogeneous and appears farther from the uniform steady-state density. However, spatial proximity alone does not determine the relaxation rate. As indicated in the inset of Fig.~\ref{fig5}(b),
the localized state has a substantially smaller overlap with the slowest Liouvillian decay mode. Since the long-time dynamics are governed by this slow mode, the reduced spectral weight naturally leads to accelerated convergence. The relaxation hierarchy is therefore dictated by the distribution of Liouvillian projections rather than by spatial structure.

\begin{figure}[t!]
	\centering
	\includegraphics[width=1\linewidth]{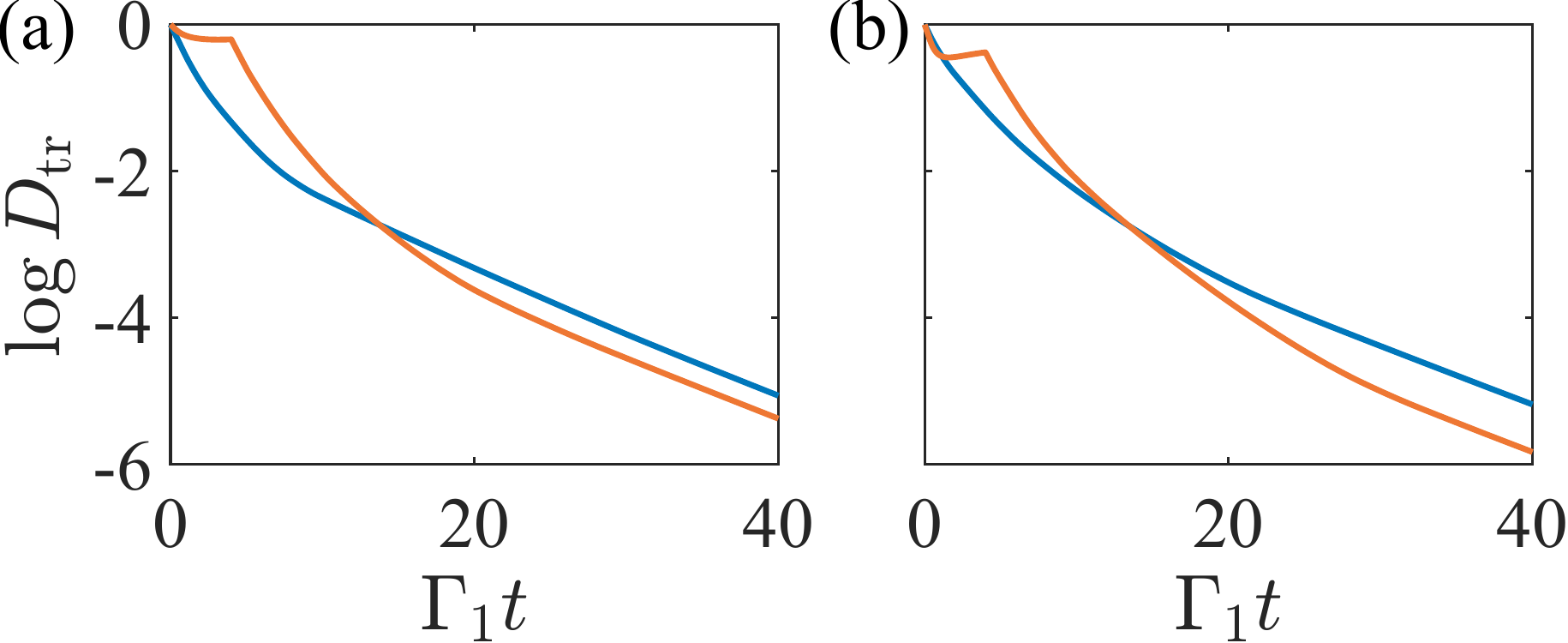}
	\caption{
		Relaxation dynamics under the two-step (orange) and one-step (blue) protocols
		in the power-law hopping model at $L=90$, $t=1$, $V=1.1$, and $a=1$.
		The trace distance $D_{\mathrm{tr}}$ from the infinite-temperature steady state
		is shown as a function of time.
		During the prethermalization stage, 
		the system is weakly coupled to an auxiliary bath 
		with coupling strength $\Gamma_2=\Gamma_1/100$ 
		for a duration equal to one-tenth of the total evolution time.
		(a) 15\text{th}  eigenstate
		($E=-1.62$, $T_{\mathrm{eff}}=3.72$),
		prethermalized at $T=1$.
		(b) 30\text{th}  eigenstate
		($E=-0.54$, $T_{\mathrm{eff}}=8.63$),
		prethermalized at $T=2$.
		In both cases, the two-step protocol
		exhibits a crossover with the one-step dynamics
		and achieves faster convergence at late times,
		signaling the PME.
	}
	\label{fig6}
\end{figure}

We next examine the robustness of the acceleration protocol within the same model. We consider two additional eigenstates in the $P_1$ phase: the 15\text{th} state ($E=-1.62$, $T_{\mathrm{eff}}=3.72$)
and the 30\text{th} state ($E=-0.54$, $T_{\mathrm{eff}}=8.63$).
In the two-step protocol, the system is first weakly coupled to an auxiliary thermal bath with $\Gamma_2=\Gamma_1/100$, at temperatures $T=1$ and $T=2$, respectively, both lower than the corresponding effective temperatures. The subsequent evolution proceeds under the primary dissipator with rate $\Gamma_1$. As shown in the Fig.~\ref{fig6},
both initial states display the characteristic crossing behavior
associated with the PME. While the direct (one-step) protocol exhibits a faster initial decay, the two-step protocol progressively accelerates and eventually overtakes the direct trajectory, after which no further crossings occur. This behavior indicates that the intermediate evolution reshapes the spectral composition of the state,
reducing its projection onto the slow Liouvillian modes that dominate long-time relaxation. As a consequence, the composite protocol achieves a shorter total relaxation time.

From a thermodynamic viewpoint, this realization may be interpreted as a temperature-inverted PME, where an initial cooling stage counterintuitively facilitates subsequent equilibration. This initial cooling reduces the projection onto slow Liouvillian modes and shortens the overall equilibration time, realizing a temperature-inverted PME in a system with mobility edges and long-range hopping. The consistency of this mechanism across both
nearest-neighbor and long-range quasiperiodic models indicates that the effect is governed by a general dynamical principle rather than by specific microscopic details.

\section{Conclusion and outlook}
\label{Conclusion}

In this work, we have demonstrated the emergence of the Pontus quantum Mpemba effect in one-dimensional quasiperiodic systems subject to Markovian dephasing. By analyzing both a nearest-neighbor tight-binding chain with a quasiperiodic mosaic potential and its long-range hopping counterpart~\cite{Mosaic1,LongRange1}, we show that an appropriately designed two-step Pontus protocol yields systematically faster convergence toward the infinite-temperature steady state than direct evolution from the same initial condition. This acceleration is observed for initial states with distinct localization characteristics and persists in the presence of power-law long-range couplings, indicating that the effect is not contingent upon specific locality constraints or fine-tuned spectral details of a particular Hamiltonian realization.

The physical origin of this acceleration lies in the spectral structure of the Liouvillian superoperator governing the dissipative dynamics. The intermediate state generated during the first stage of the protocol redistributes the spectral decomposition of the density matrix and reduces its projection onto the slowest-decaying Liouvillian eigenmodes that dominate late-time relaxation. Consequently, the effective relaxation timescale becomes controlled by faster decay channels, leading to a genuine inversion of relaxation efficiency relative to the direct protocol. Our analysis indicates that this mechanism operates largely independently of whether the underlying Hamiltonian eigenstates are localized or extended, underscoring that Pontus acceleration is fundamentally a property of Liouvillian spectral geometry rather than a static feature of the closed-system eigenstructure.

Conceptually, our results extend Mpemba-like phenomena to quasiperiodic open systems within a protocol-based framework. In contrast to conventional quantum Mpemba effects, which compare distinct initial states evolving under identical dynamics and may rely on particular symmetries or critical spectral structures~\cite{QME_Exp3,OpenQME1}--the Pontus scheme achieves acceleration through controlled intermediate-state engineering starting from a common initial configuration~\cite{QME27,QME28}. Given that both quasiperiodic potentials and engineered dephasing can be realized in existing quantum simulation platforms~\cite{Exp1,Exp2,Exp3,Exp4,QME_Exp4}, experimental verification appears within reach. More broadly, our findings suggest that Liouvillian spectral engineering provides a versatile route for controlling dissipative relaxation in complex quantum systems. Extensions to interacting quasiperiodic lattices, non-Markovian environments~\cite{OpenQME6}, and alternative dissipative architectures may further elucidate how relaxation pathways can be systematically optimized in open quantum many-body settings.

\section*{Acknowledgments}
The work is supported by the National Natural Science Foundation of China (Grant No. 12304290 and No. 12505017), and Beijing National Laboratory for Condensed Matter Physics (2025BNLCMPKF017). LP also acknowledges support from the Fundamental Research Funds for the Central Universities. \\

\bibliography{Ref}

\begin{thebibliography}{95}%
\makeatletter
\providecommand \@ifxundefined [1]{%
 \@ifx{#1\undefined}
}%
\providecommand \@ifnum [1]{%
 \ifnum #1\expandafter \@firstoftwo
 \else \expandafter \@secondoftwo
 \fi
}%
\providecommand \@ifx [1]{%
 \ifx #1\expandafter \@firstoftwo
 \else \expandafter \@secondoftwo
 \fi
}%
\providecommand \natexlab [1]{#1}%
\providecommand \enquote  [1]{``#1''}%
\providecommand \bibnamefont  [1]{#1}%
\providecommand \bibfnamefont [1]{#1}%
\providecommand \citenamefont [1]{#1}%
\providecommand \href@noop [0]{\@secondoftwo}%
\providecommand \href [0]{\begingroup \@sanitize@url \@href}%
\providecommand \@href[1]{\@@startlink{#1}\@@href}%
\providecommand \@@href[1]{\endgroup#1\@@endlink}%
\providecommand \@sanitize@url [0]{\catcode `\\12\catcode `\$12\catcode
  `\&12\catcode `\#12\catcode `\^12\catcode `\_12\catcode `\%12\relax}%
\providecommand \@@startlink[1]{}%
\providecommand \@@endlink[0]{}%
\providecommand \url  [0]{\begingroup\@sanitize@url \@url }%
\providecommand \@url [1]{\endgroup\@href {#1}{\urlprefix }}%
\providecommand \urlprefix  [0]{URL }%
\providecommand \Eprint [0]{\href }%
\providecommand \doibase [0]{https://doi.org/}%
\providecommand \selectlanguage [0]{\@gobble}%
\providecommand \bibinfo  [0]{\@secondoftwo}%
\providecommand \bibfield  [0]{\@secondoftwo}%
\providecommand \translation [1]{[#1]}%
\providecommand \BibitemOpen [0]{}%
\providecommand \bibitemStop [0]{}%
\providecommand \bibitemNoStop [0]{.\EOS\space}%
\providecommand \EOS [0]{\spacefactor3000\relax}%
\providecommand \BibitemShut  [1]{\csname bibitem#1\endcsname}%
\let\auto@bib@innerbib\@empty
\bibitem [{\citenamefont {Mpemba}\ and\ \citenamefont {Osborne}(1969)}]{ME}%
  \BibitemOpen
  \bibfield  {author} {\bibinfo {author} {\bibfnamefont {E.~B.}\ \bibnamefont
  {Mpemba}}\ and\ \bibinfo {author} {\bibfnamefont {D.~G.}\ \bibnamefont
  {Osborne}},\ }\bibfield  {title} {\bibinfo {title} {Cool?},\ }\href
  {https://doi.org/10.1088/0031-9120/4/3/312} {\bibfield  {journal} {\bibinfo
  {journal} {Phys. Educ.}\ }\textbf {\bibinfo {volume} {4}},\ \bibinfo {pages}
  {172} (\bibinfo {year} {1969})}\BibitemShut {NoStop}%
\bibitem [{\citenamefont {Lasanta}\ \emph {et~al.}(2017)\citenamefont
  {Lasanta}, \citenamefont {Vega~Reyes}, \citenamefont {Prados},\ and\
  \citenamefont {Santos}}]{ME_classical9}%
  \BibitemOpen
  \bibfield  {author} {\bibinfo {author} {\bibfnamefont {A.}~\bibnamefont
  {Lasanta}}, \bibinfo {author} {\bibfnamefont {F.}~\bibnamefont {Vega~Reyes}},
  \bibinfo {author} {\bibfnamefont {A.}~\bibnamefont {Prados}},\ and\ \bibinfo
  {author} {\bibfnamefont {A.}~\bibnamefont {Santos}},\ }\bibfield  {title}
  {\bibinfo {title} {When the hotter cools more quickly: Mpemba effect in
  granular fluids},\ }\href {https://doi.org/10.1103/PhysRevLett.119.148001}
  {\bibfield  {journal} {\bibinfo  {journal} {Phys. Rev. Lett.}\ }\textbf
  {\bibinfo {volume} {119}},\ \bibinfo {pages} {148001} (\bibinfo {year}
  {2017})}\BibitemShut {NoStop}%
\bibitem [{\citenamefont {Kumar}\ and\ \citenamefont
  {Bechhoefer}(2020)}]{ME_classical5}%
  \BibitemOpen
  \bibfield  {author} {\bibinfo {author} {\bibfnamefont {A.}~\bibnamefont
  {Kumar}}\ and\ \bibinfo {author} {\bibfnamefont {J.}~\bibnamefont
  {Bechhoefer}},\ }\bibfield  {title} {\bibinfo {title} {Exponentially faster
  cooling in a colloidal system},\ }\href
  {https://doi.org/10.1038/s41586-020-2560-x} {\bibfield  {journal} {\bibinfo
  {journal} {Nature (London)}\ }\textbf {\bibinfo {volume} {584}},\ \bibinfo
  {pages} {64} (\bibinfo {year} {2020})}\BibitemShut {NoStop}%
\bibitem [{\citenamefont {Liu}\ \emph {et~al.}(2023)\citenamefont {Liu},
  \citenamefont {Li}, \citenamefont {Liu}, \citenamefont {Hamley},\ and\
  \citenamefont {Jiang}}]{ME_classical7}%
  \BibitemOpen
  \bibfield  {author} {\bibinfo {author} {\bibfnamefont {J.}~\bibnamefont
  {Liu}}, \bibinfo {author} {\bibfnamefont {J.}~\bibnamefont {Li}}, \bibinfo
  {author} {\bibfnamefont {B.}~\bibnamefont {Liu}}, \bibinfo {author}
  {\bibfnamefont {I.~W.}\ \bibnamefont {Hamley}},\ and\ \bibinfo {author}
  {\bibfnamefont {S.}~\bibnamefont {Jiang}},\ }\bibfield  {title} {\bibinfo
  {title} {Mpemba effect in crystallization of polybutene-1},\ }\href
  {https://doi.org/10.1039/D3SM00309D} {\bibfield  {journal} {\bibinfo
  {journal} {Soft Matter}\ }\textbf {\bibinfo {volume} {19}},\ \bibinfo {pages}
  {3337} (\bibinfo {year} {2023})}\BibitemShut {NoStop}%
\bibitem [{\citenamefont {Chorazewski}\ \emph {et~al.}(2024)\citenamefont
  {Chorazewski}, \citenamefont {Wasiak}, \citenamefont {Sychev}, \citenamefont
  {Korotkovskii},\ and\ \citenamefont {Postnikov}}]{ME_classical8}%
  \BibitemOpen
  \bibfield  {author} {\bibinfo {author} {\bibfnamefont {M.}~\bibnamefont
  {Chorazewski}}, \bibinfo {author} {\bibfnamefont {M.}~\bibnamefont {Wasiak}},
  \bibinfo {author} {\bibfnamefont {A.~V.}\ \bibnamefont {Sychev}}, \bibinfo
  {author} {\bibfnamefont {V.~I.}\ \bibnamefont {Korotkovskii}},\ and\ \bibinfo
  {author} {\bibfnamefont {E.~B.}\ \bibnamefont {Postnikov}},\ }\bibfield
  {title} {\bibinfo {title} {The curious case of 1-ethylpyridinium triflate:
  Ionic liquid exhibiting the mpemba effect},\ }\href
  {https://doi.org/10.1007/s10953-023-01268-1} {\bibfield  {journal} {\bibinfo
  {journal} {J. Solution Chem.}\ }\textbf {\bibinfo {volume} {53}},\ \bibinfo
  {pages} {80} (\bibinfo {year} {2024})}\BibitemShut {NoStop}%
\bibitem [{\citenamefont {Jeng}(2006)}]{ME_classical1}%
  \BibitemOpen
  \bibfield  {author} {\bibinfo {author} {\bibfnamefont {M.}~\bibnamefont
  {Jeng}},\ }\bibfield  {title} {\bibinfo {title} {The mpemba effect: When can
  hot water freeze faster than cold?},\ }\href
  {https://doi.org/10.1119/1.2186331} {\bibfield  {journal} {\bibinfo
  {journal} {Am. J. Phys.}\ }\textbf {\bibinfo {volume} {74}},\ \bibinfo
  {pages} {514} (\bibinfo {year} {2006})}\BibitemShut {NoStop}%
\bibitem [{\citenamefont {Klich}\ \emph {et~al.}(2019)\citenamefont {Klich},
  \citenamefont {Raz}, \citenamefont {Hirschberg},\ and\ \citenamefont
  {Vucelja}}]{ME_classical2}%
  \BibitemOpen
  \bibfield  {author} {\bibinfo {author} {\bibfnamefont {I.}~\bibnamefont
  {Klich}}, \bibinfo {author} {\bibfnamefont {O.}~\bibnamefont {Raz}}, \bibinfo
  {author} {\bibfnamefont {O.}~\bibnamefont {Hirschberg}},\ and\ \bibinfo
  {author} {\bibfnamefont {M.}~\bibnamefont {Vucelja}},\ }\bibfield  {title}
  {\bibinfo {title} {Mpemba index and anomalous relaxation},\ }\href
  {https://doi.org/10.1103/PhysRevX.9.021060} {\bibfield  {journal} {\bibinfo
  {journal} {Phys. Rev. X}\ }\textbf {\bibinfo {volume} {9}},\ \bibinfo {pages}
  {021060} (\bibinfo {year} {2019})}\BibitemShut {NoStop}%
\bibitem [{\citenamefont {Ahn}\ \emph {et~al.}(2016)\citenamefont {Ahn},
  \citenamefont {Kang}, \citenamefont {Koh},\ and\ \citenamefont
  {Lee}}]{ME_classical3}%
  \BibitemOpen
  \bibfield  {author} {\bibinfo {author} {\bibfnamefont {Y.-H.}\ \bibnamefont
  {Ahn}}, \bibinfo {author} {\bibfnamefont {H.}~\bibnamefont {Kang}}, \bibinfo
  {author} {\bibfnamefont {D.-Y.}\ \bibnamefont {Koh}},\ and\ \bibinfo {author}
  {\bibfnamefont {H.}~\bibnamefont {Lee}},\ }\bibfield  {title} {\bibinfo
  {title} {Experimental verifications of mpemba-like behaviors of clathrate
  hydrates},\ }\href {https://doi.org/10.1007/s11814-016-0029-2} {\bibfield
  {journal} {\bibinfo  {journal} {Korean J. Chem. Eng.}\ }\textbf {\bibinfo
  {volume} {33}},\ \bibinfo {pages} {1903} (\bibinfo {year}
  {2016})}\BibitemShut {NoStop}%
\bibitem [{\citenamefont {Chaddah}\ \emph {et~al.}(2010)\citenamefont
  {Chaddah}, \citenamefont {Dash}, \citenamefont {Kumar},\ and\ \citenamefont
  {Banerjee}}]{ME_classical4}%
  \BibitemOpen
  \bibfield  {author} {\bibinfo {author} {\bibfnamefont {P.}~\bibnamefont
  {Chaddah}}, \bibinfo {author} {\bibfnamefont {S.}~\bibnamefont {Dash}},
  \bibinfo {author} {\bibfnamefont {K.}~\bibnamefont {Kumar}},\ and\ \bibinfo
  {author} {\bibfnamefont {A.}~\bibnamefont {Banerjee}},\ }\bibfield  {title}
  {\bibinfo {title} {Overtaking while approaching equilibrium},\ }\href@noop {}
  {\bibfield  {journal} {\bibinfo  {journal} {arXiv preprint}\ } (\bibinfo
  {year} {2010})},\ \Eprint {https://arxiv.org/abs/1011.3598} {arXiv:1011.3598}
  \BibitemShut {NoStop}%
\bibitem [{\citenamefont {Hu}\ \emph {et~al.}(2018)\citenamefont {Hu},
  \citenamefont {Li}, \citenamefont {Huang}, \citenamefont {Li}, \citenamefont
  {Luo}, \citenamefont {Chen}, \citenamefont {Jiang},\ and\ \citenamefont
  {An}}]{ME_classical6}%
  \BibitemOpen
  \bibfield  {author} {\bibinfo {author} {\bibfnamefont {C.}~\bibnamefont
  {Hu}}, \bibinfo {author} {\bibfnamefont {J.}~\bibnamefont {Li}}, \bibinfo
  {author} {\bibfnamefont {S.}~\bibnamefont {Huang}}, \bibinfo {author}
  {\bibfnamefont {H.}~\bibnamefont {Li}}, \bibinfo {author} {\bibfnamefont
  {C.}~\bibnamefont {Luo}}, \bibinfo {author} {\bibfnamefont {J.}~\bibnamefont
  {Chen}}, \bibinfo {author} {\bibfnamefont {S.}~\bibnamefont {Jiang}},\ and\
  \bibinfo {author} {\bibfnamefont {L.}~\bibnamefont {An}},\ }\bibfield
  {title} {\bibinfo {title} {Conformation directed mpemba effect on polylactide
  crystallization},\ }\href {https://doi.org/10.1021/acs.cgd.8b01250}
  {\bibfield  {journal} {\bibinfo  {journal} {Cryst. Growth Des.}\ }\textbf
  {\bibinfo {volume} {18}},\ \bibinfo {pages} {5757} (\bibinfo {year}
  {2018})}\BibitemShut {NoStop}%
\bibitem [{\citenamefont {Lu}\ and\ \citenamefont {Raz}(2017)}]{Inverse_ME1}%
  \BibitemOpen
  \bibfield  {author} {\bibinfo {author} {\bibfnamefont {Z.}~\bibnamefont
  {Lu}}\ and\ \bibinfo {author} {\bibfnamefont {O.}~\bibnamefont {Raz}},\
  }\bibfield  {title} {\bibinfo {title} {Nonequilibrium thermodynamics of the
  markovian mpemba effect and its inverse},\ }\href
  {https://doi.org/10.1073/pnas.1701264114} {\bibfield  {journal} {\bibinfo
  {journal} {Proc. Natl. Acad. Sci. U.S.A.}\ }\textbf {\bibinfo {volume}
  {114}},\ \bibinfo {pages} {5083} (\bibinfo {year} {2017})}\BibitemShut
  {NoStop}%
\bibitem [{\citenamefont {Shapira}\ \emph {et~al.}(2024)\citenamefont
  {Shapira}, \citenamefont {Shapira}, \citenamefont {Markov}, \citenamefont
  {Teza}, \citenamefont {Akerman}, \citenamefont {Raz},\ and\ \citenamefont
  {Ozeri}}]{Inverse_ME2}%
  \BibitemOpen
  \bibfield  {author} {\bibinfo {author} {\bibfnamefont {S.~A.}\ \bibnamefont
  {Shapira}}, \bibinfo {author} {\bibfnamefont {Y.}~\bibnamefont {Shapira}},
  \bibinfo {author} {\bibfnamefont {J.}~\bibnamefont {Markov}}, \bibinfo
  {author} {\bibfnamefont {G.}~\bibnamefont {Teza}}, \bibinfo {author}
  {\bibfnamefont {N.}~\bibnamefont {Akerman}}, \bibinfo {author} {\bibfnamefont
  {O.}~\bibnamefont {Raz}},\ and\ \bibinfo {author} {\bibfnamefont
  {R.}~\bibnamefont {Ozeri}},\ }\bibfield  {title} {\bibinfo {title} {Inverse
  mpemba effect demonstrated on a single trapped ion qubit},\ }\href
  {https://doi.org/10.1103/PhysRevLett.133.010403} {\bibfield  {journal}
  {\bibinfo  {journal} {Phys. Rev. Lett.}\ }\textbf {\bibinfo {volume} {133}},\
  \bibinfo {pages} {010403} (\bibinfo {year} {2024})}\BibitemShut {NoStop}%
\bibitem [{\citenamefont {Kumar}\ \emph {et~al.}(2022)\citenamefont {Kumar},
  \citenamefont {Chétrite},\ and\ \citenamefont {Bechhoefer}}]{Inverse_ME3}%
  \BibitemOpen
  \bibfield  {author} {\bibinfo {author} {\bibfnamefont {A.}~\bibnamefont
  {Kumar}}, \bibinfo {author} {\bibfnamefont {R.}~\bibnamefont {Chétrite}},\
  and\ \bibinfo {author} {\bibfnamefont {J.}~\bibnamefont {Bechhoefer}},\
  }\bibfield  {title} {\bibinfo {title} {Anomalous heating in a colloidal
  system},\ }\href {https://doi.org/10.1073/pnas.2118484119} {\bibfield
  {journal} {\bibinfo  {journal} {Proc. Natl. Acad. Sci. U.S.A.}\ }\textbf
  {\bibinfo {volume} {119}},\ \bibinfo {pages} {e2118484119} (\bibinfo {year}
  {2022})}\BibitemShut {NoStop}%
\bibitem [{\citenamefont {Joshi}\ \emph {et~al.}(2024)\citenamefont {Joshi},
  \citenamefont {Franke}, \citenamefont {Rath}, \citenamefont {Ares},
  \citenamefont {Murciano}, \citenamefont {Kranzl}, \citenamefont {Blatt},
  \citenamefont {Zoller}, \citenamefont {Vermersch}, \citenamefont {Calabrese},
  \citenamefont {Roos},\ and\ \citenamefont {Joshi}}]{QME_Exp1}%
  \BibitemOpen
  \bibfield  {author} {\bibinfo {author} {\bibfnamefont {L.~K.}\ \bibnamefont
  {Joshi}}, \bibinfo {author} {\bibfnamefont {J.}~\bibnamefont {Franke}},
  \bibinfo {author} {\bibfnamefont {A.}~\bibnamefont {Rath}}, \bibinfo {author}
  {\bibfnamefont {F.}~\bibnamefont {Ares}}, \bibinfo {author} {\bibfnamefont
  {S.}~\bibnamefont {Murciano}}, \bibinfo {author} {\bibfnamefont
  {F.}~\bibnamefont {Kranzl}}, \bibinfo {author} {\bibfnamefont
  {R.}~\bibnamefont {Blatt}}, \bibinfo {author} {\bibfnamefont
  {P.}~\bibnamefont {Zoller}}, \bibinfo {author} {\bibfnamefont
  {B.}~\bibnamefont {Vermersch}}, \bibinfo {author} {\bibfnamefont
  {P.}~\bibnamefont {Calabrese}}, \bibinfo {author} {\bibfnamefont {C.~F.}\
  \bibnamefont {Roos}},\ and\ \bibinfo {author} {\bibfnamefont {M.~K.}\
  \bibnamefont {Joshi}},\ }\bibfield  {title} {\bibinfo {title} {Observing the
  quantum mpemba effect in quantum simulations},\ }\href
  {https://doi.org/10.1103/PhysRevLett.133.010402} {\bibfield  {journal}
  {\bibinfo  {journal} {Phys. Rev. Lett.}\ }\textbf {\bibinfo {volume} {133}},\
  \bibinfo {pages} {010402} (\bibinfo {year} {2024})}\BibitemShut {NoStop}%
\bibitem [{\citenamefont {Zhang}\ \emph {et~al.}(2025)\citenamefont {Zhang},
  \citenamefont {Xia}, \citenamefont {Wu}, \citenamefont {Chen}, \citenamefont
  {Zhang}, \citenamefont {Xie}, \citenamefont {Su}, \citenamefont {Wu},
  \citenamefont {Qiu}, \citenamefont {Chen}, \citenamefont {Li}, \citenamefont
  {Jing},\ and\ \citenamefont {Zhou}}]{QME_Exp2}%
  \BibitemOpen
  \bibfield  {author} {\bibinfo {author} {\bibfnamefont {J.}~\bibnamefont
  {Zhang}}, \bibinfo {author} {\bibfnamefont {G.}~\bibnamefont {Xia}}, \bibinfo
  {author} {\bibfnamefont {C.-W.}\ \bibnamefont {Wu}}, \bibinfo {author}
  {\bibfnamefont {T.}~\bibnamefont {Chen}}, \bibinfo {author} {\bibfnamefont
  {Q.}~\bibnamefont {Zhang}}, \bibinfo {author} {\bibfnamefont
  {Y.}~\bibnamefont {Xie}}, \bibinfo {author} {\bibfnamefont {W.-B.}\
  \bibnamefont {Su}}, \bibinfo {author} {\bibfnamefont {W.}~\bibnamefont {Wu}},
  \bibinfo {author} {\bibfnamefont {C.-W.}\ \bibnamefont {Qiu}}, \bibinfo
  {author} {\bibfnamefont {P.-X.}\ \bibnamefont {Chen}}, \bibinfo {author}
  {\bibfnamefont {W.}~\bibnamefont {Li}}, \bibinfo {author} {\bibfnamefont
  {H.}~\bibnamefont {Jing}},\ and\ \bibinfo {author} {\bibfnamefont {Y.-L.}\
  \bibnamefont {Zhou}},\ }\bibfield  {title} {\bibinfo {title} {Observation of
  quantum strong mpemba effect},\ }\href
  {https://doi.org/10.1038/s41467-024-54303-0} {\bibfield  {journal} {\bibinfo
  {journal} {Nat. Commun.}\ }\textbf {\bibinfo {volume} {16}},\ \bibinfo
  {pages} {301} (\bibinfo {year} {2025})}\BibitemShut {NoStop}%
\bibitem [{\citenamefont {Yu}\ \emph {et~al.}(2025{\natexlab{a}})\citenamefont
  {Yu}, \citenamefont {Jin}, \citenamefont {Zhang}, \citenamefont {Xu},\ and\
  \citenamefont {Fan}}]{QME_Exp3}%
  \BibitemOpen
  \bibfield  {author} {\bibinfo {author} {\bibfnamefont {Y.-H.}\ \bibnamefont
  {Yu}}, \bibinfo {author} {\bibfnamefont {T.-R.}\ \bibnamefont {Jin}},
  \bibinfo {author} {\bibfnamefont {L.}~\bibnamefont {Zhang}}, \bibinfo
  {author} {\bibfnamefont {K.}~\bibnamefont {Xu}},\ and\ \bibinfo {author}
  {\bibfnamefont {H.}~\bibnamefont {Fan}},\ }\bibfield  {title} {\bibinfo
  {title} {Tuning the quantum mpemba effect in an isolated system by
  initial-state engineering},\ }\href {https://doi.org/10.1103/yzjd-pk8h}
  {\bibfield  {journal} {\bibinfo  {journal} {Phys. Rev. B}\ }\textbf {\bibinfo
  {volume} {112}},\ \bibinfo {pages} {094315} (\bibinfo {year}
  {2025}{\natexlab{a}})}\BibitemShut {NoStop}%
\bibitem [{\citenamefont {Xu}\ \emph {et~al.}(2025{\natexlab{a}})\citenamefont
  {Xu}, \citenamefont {Fang}, \citenamefont {Chen}, \citenamefont {Wang},
  \citenamefont {Ge}, \citenamefont {Shi}, \citenamefont {Liu}, \citenamefont
  {Deng}, \citenamefont {Zhao}, \citenamefont {Liu}, \citenamefont {Li},
  \citenamefont {Li}, \citenamefont {Wang}, \citenamefont {Liang},
  \citenamefont {Feng}, \citenamefont {Guo}, \citenamefont {Gu}, \citenamefont
  {He}, \citenamefont {Liu}, \citenamefont {Mei}, \citenamefont {Xiao},
  \citenamefont {Yan}, \citenamefont {Yu}, \citenamefont {Yuan}, \citenamefont
  {Zhang}, \citenamefont {Wang}, \citenamefont {Liu}, \citenamefont {Song},
  \citenamefont {Tian}, \citenamefont {Zhang}, \citenamefont {Zhang},
  \citenamefont {Huang}, \citenamefont {Xiang}, \citenamefont {Zheng},
  \citenamefont {Xu},\ and\ \citenamefont {Fan}}]{QME_Exp4}%
  \BibitemOpen
  \bibfield  {author} {\bibinfo {author} {\bibfnamefont {Y.}~\bibnamefont
  {Xu}}, \bibinfo {author} {\bibfnamefont {C.-P.}\ \bibnamefont {Fang}},
  \bibinfo {author} {\bibfnamefont {B.-J.}\ \bibnamefont {Chen}}, \bibinfo
  {author} {\bibfnamefont {M.-C.}\ \bibnamefont {Wang}}, \bibinfo {author}
  {\bibfnamefont {Z.-Y.}\ \bibnamefont {Ge}}, \bibinfo {author} {\bibfnamefont
  {Y.-H.}\ \bibnamefont {Shi}}, \bibinfo {author} {\bibfnamefont
  {Y.}~\bibnamefont {Liu}}, \bibinfo {author} {\bibfnamefont {C.-L.}\
  \bibnamefont {Deng}}, \bibinfo {author} {\bibfnamefont {K.}~\bibnamefont
  {Zhao}}, \bibinfo {author} {\bibfnamefont {Z.-H.}\ \bibnamefont {Liu}},
  \bibinfo {author} {\bibfnamefont {T.-M.}\ \bibnamefont {Li}}, \bibinfo
  {author} {\bibfnamefont {H.}~\bibnamefont {Li}}, \bibinfo {author}
  {\bibfnamefont {Z.}~\bibnamefont {Wang}}, \bibinfo {author} {\bibfnamefont
  {G.-H.}\ \bibnamefont {Liang}}, \bibinfo {author} {\bibfnamefont
  {D.}~\bibnamefont {Feng}}, \bibinfo {author} {\bibfnamefont {X.}~\bibnamefont
  {Guo}}, \bibinfo {author} {\bibfnamefont {X.-Y.}\ \bibnamefont {Gu}},
  \bibinfo {author} {\bibfnamefont {Y.}~\bibnamefont {He}}, \bibinfo {author}
  {\bibfnamefont {H.-T.}\ \bibnamefont {Liu}}, \bibinfo {author} {\bibfnamefont
  {Z.-Y.}\ \bibnamefont {Mei}}, \bibinfo {author} {\bibfnamefont
  {Y.}~\bibnamefont {Xiao}}, \bibinfo {author} {\bibfnamefont {Y.}~\bibnamefont
  {Yan}}, \bibinfo {author} {\bibfnamefont {Y.-H.}\ \bibnamefont {Yu}},
  \bibinfo {author} {\bibfnamefont {W.-P.}\ \bibnamefont {Yuan}}, \bibinfo
  {author} {\bibfnamefont {J.-C.}\ \bibnamefont {Zhang}}, \bibinfo {author}
  {\bibfnamefont {Z.-A.}\ \bibnamefont {Wang}}, \bibinfo {author}
  {\bibfnamefont {G.}~\bibnamefont {Liu}}, \bibinfo {author} {\bibfnamefont
  {X.}~\bibnamefont {Song}}, \bibinfo {author} {\bibfnamefont {Y.}~\bibnamefont
  {Tian}}, \bibinfo {author} {\bibfnamefont {Y.-R.}\ \bibnamefont {Zhang}},
  \bibinfo {author} {\bibfnamefont {S.-X.}\ \bibnamefont {Zhang}}, \bibinfo
  {author} {\bibfnamefont {K.}~\bibnamefont {Huang}}, \bibinfo {author}
  {\bibfnamefont {Z.}~\bibnamefont {Xiang}}, \bibinfo {author} {\bibfnamefont
  {D.}~\bibnamefont {Zheng}}, \bibinfo {author} {\bibfnamefont
  {K.}~\bibnamefont {Xu}},\ and\ \bibinfo {author} {\bibfnamefont
  {H.}~\bibnamefont {Fan}},\ }\bibfield  {title} {\bibinfo {title} {Observation
  and modulation of the quantum mpemba effect on a superconducting quantum
  processor},\ }\href@noop {} {\bibfield  {journal} {\bibinfo  {journal} {arXiv
  preprint}\ } (\bibinfo {year} {2025}{\natexlab{a}})},\ \Eprint
  {https://arxiv.org/abs/2508.07707} {arXiv:2508.07707} \BibitemShut {NoStop}%
\bibitem [{\citenamefont {Ares}\ \emph
  {et~al.}(2025{\natexlab{a}})\citenamefont {Ares}, \citenamefont {Calabrese},\
  and\ \citenamefont {Murciano}}]{QME1}%
  \BibitemOpen
  \bibfield  {author} {\bibinfo {author} {\bibfnamefont {F.}~\bibnamefont
  {Ares}}, \bibinfo {author} {\bibfnamefont {P.}~\bibnamefont {Calabrese}},\
  and\ \bibinfo {author} {\bibfnamefont {S.}~\bibnamefont {Murciano}},\
  }\bibfield  {title} {\bibinfo {title} {The quantum mpemba effects},\ }\href
  {https://doi.org/10.1038/s42254-025-00838-0} {\bibfield  {journal} {\bibinfo
  {journal} {Nat. Rev. Phys.}\ }\textbf {\bibinfo {volume} {7}},\ \bibinfo
  {pages} {451} (\bibinfo {year} {2025}{\natexlab{a}})}\BibitemShut {NoStop}%
\bibitem [{\citenamefont {Yu}\ \emph {et~al.}(2025{\natexlab{b}})\citenamefont
  {Yu}, \citenamefont {Liu},\ and\ \citenamefont {Zhang}}]{QME101}%
  \BibitemOpen
  \bibfield  {author} {\bibinfo {author} {\bibfnamefont {H.}~\bibnamefont
  {Yu}}, \bibinfo {author} {\bibfnamefont {S.}~\bibnamefont {Liu}},\ and\
  \bibinfo {author} {\bibfnamefont {S.-X.}\ \bibnamefont {Zhang}},\ }\bibfield
  {title} {\bibinfo {title} {Quantum mpemba effects from symmetry
  perspectives},\ }\href {https://doi.org/10.1007/s43673-025-00157-7}
  {\bibfield  {journal} {\bibinfo  {journal} {AAPPS Bull.}\ }\textbf {\bibinfo
  {volume} {35}},\ \bibinfo {pages} {17} (\bibinfo {year}
  {2025}{\natexlab{b}})}\BibitemShut {NoStop}%
\bibitem [{\citenamefont {Murciano}\ \emph {et~al.}(2024)\citenamefont
  {Murciano}, \citenamefont {Ares}, \citenamefont {Klich},\ and\ \citenamefont
  {Calabrese}}]{QME2}%
  \BibitemOpen
  \bibfield  {author} {\bibinfo {author} {\bibfnamefont {S.}~\bibnamefont
  {Murciano}}, \bibinfo {author} {\bibfnamefont {F.}~\bibnamefont {Ares}},
  \bibinfo {author} {\bibfnamefont {I.}~\bibnamefont {Klich}},\ and\ \bibinfo
  {author} {\bibfnamefont {P.}~\bibnamefont {Calabrese}},\ }\bibfield  {title}
  {\bibinfo {title} {Entanglement asymmetry and quantum mpemba effect in the xy
  spin chain},\ }\href {https://doi.org/10.1088/1742-5468/ad17b4} {\bibfield
  {journal} {\bibinfo  {journal} {J. Stat. Mech.}\ }\textbf {\bibinfo {volume}
  {2024}},\ \bibinfo {pages} {013103} (\bibinfo {year} {2024})}\BibitemShut
  {NoStop}%
\bibitem [{\citenamefont {Chalas}\ \emph {et~al.}(2024)\citenamefont {Chalas},
  \citenamefont {Ares}, \citenamefont {Rylands},\ and\ \citenamefont
  {Calabrese}}]{QME3}%
  \BibitemOpen
  \bibfield  {author} {\bibinfo {author} {\bibfnamefont {K.}~\bibnamefont
  {Chalas}}, \bibinfo {author} {\bibfnamefont {F.}~\bibnamefont {Ares}},
  \bibinfo {author} {\bibfnamefont {C.}~\bibnamefont {Rylands}},\ and\ \bibinfo
  {author} {\bibfnamefont {P.}~\bibnamefont {Calabrese}},\ }\bibfield  {title}
  {\bibinfo {title} {Multiple crossing during dynamical symmetry restoration
  and implications for the quantum mpemba effect},\ }\href
  {https://doi.org/10.1088/1742-5468/ad769c} {\bibfield  {journal} {\bibinfo
  {journal} {J. Stat. Mech.}\ }\textbf {\bibinfo {volume} {2024}},\ \bibinfo
  {pages} {103101} (\bibinfo {year} {2024})}\BibitemShut {NoStop}%
\bibitem [{\citenamefont {Rylands}\ \emph {et~al.}(2024)\citenamefont
  {Rylands}, \citenamefont {Klobas}, \citenamefont {Ares}, \citenamefont
  {Calabrese}, \citenamefont {Murciano},\ and\ \citenamefont {Bertini}}]{QME4}%
  \BibitemOpen
  \bibfield  {author} {\bibinfo {author} {\bibfnamefont {C.}~\bibnamefont
  {Rylands}}, \bibinfo {author} {\bibfnamefont {K.}~\bibnamefont {Klobas}},
  \bibinfo {author} {\bibfnamefont {F.}~\bibnamefont {Ares}}, \bibinfo {author}
  {\bibfnamefont {P.}~\bibnamefont {Calabrese}}, \bibinfo {author}
  {\bibfnamefont {S.}~\bibnamefont {Murciano}},\ and\ \bibinfo {author}
  {\bibfnamefont {B.}~\bibnamefont {Bertini}},\ }\bibfield  {title} {\bibinfo
  {title} {Microscopic origin of the quantum mpemba effect in integrable
  systems},\ }\href {https://doi.org/10.1103/PhysRevLett.133.010401} {\bibfield
   {journal} {\bibinfo  {journal} {Phys. Rev. Lett.}\ }\textbf {\bibinfo
  {volume} {133}},\ \bibinfo {pages} {010401} (\bibinfo {year}
  {2024})}\BibitemShut {NoStop}%
\bibitem [{\citenamefont {Lastres}\ \emph {et~al.}(2025)\citenamefont
  {Lastres}, \citenamefont {Murciano}, \citenamefont {Ares},\ and\
  \citenamefont {Calabrese}}]{QME41}%
  \BibitemOpen
  \bibfield  {author} {\bibinfo {author} {\bibfnamefont {M.}~\bibnamefont
  {Lastres}}, \bibinfo {author} {\bibfnamefont {S.}~\bibnamefont {Murciano}},
  \bibinfo {author} {\bibfnamefont {F.}~\bibnamefont {Ares}},\ and\ \bibinfo
  {author} {\bibfnamefont {P.}~\bibnamefont {Calabrese}},\ }\bibfield  {title}
  {\bibinfo {title} {Entanglement asymmetry in the critical xxz spin chain},\
  }\href {https://doi.org/10.1088/1742-5468/ada497} {\bibfield  {journal}
  {\bibinfo  {journal} {J. Stat. Mech.}\ }\textbf {\bibinfo {volume} {2025}},\
  \bibinfo {pages} {013107} (\bibinfo {year} {2025})}\BibitemShut {NoStop}%
\bibitem [{\citenamefont {Liu}\ \emph {et~al.}(2025)\citenamefont {Liu},
  \citenamefont {Zhang}, \citenamefont {Yin}, \citenamefont {Zhang},\ and\
  \citenamefont {Yao}}]{QME5}%
  \BibitemOpen
  \bibfield  {author} {\bibinfo {author} {\bibfnamefont {S.}~\bibnamefont
  {Liu}}, \bibinfo {author} {\bibfnamefont {H.-K.}\ \bibnamefont {Zhang}},
  \bibinfo {author} {\bibfnamefont {S.}~\bibnamefont {Yin}}, \bibinfo {author}
  {\bibfnamefont {S.-X.}\ \bibnamefont {Zhang}},\ and\ \bibinfo {author}
  {\bibfnamefont {H.}~\bibnamefont {Yao}},\ }\bibfield  {title} {\bibinfo
  {title} {Quantum mpemba effects in many-body localization systems},\
  }\bibfield  {journal} {\bibinfo  {journal} {Science Bulletin}\ }\href
  {https://doi.org/10.1016/j.scib.2025.10.017} {10.1016/j.scib.2025.10.017}
  (\bibinfo {year} {2025})\BibitemShut {NoStop}%
\bibitem [{\citenamefont {Liu}\ \emph {et~al.}(2024{\natexlab{a}})\citenamefont
  {Liu}, \citenamefont {Zhang}, \citenamefont {Yin},\ and\ \citenamefont
  {Zhang}}]{QME8}%
  \BibitemOpen
  \bibfield  {author} {\bibinfo {author} {\bibfnamefont {S.}~\bibnamefont
  {Liu}}, \bibinfo {author} {\bibfnamefont {H.-K.}\ \bibnamefont {Zhang}},
  \bibinfo {author} {\bibfnamefont {S.}~\bibnamefont {Yin}},\ and\ \bibinfo
  {author} {\bibfnamefont {S.-X.}\ \bibnamefont {Zhang}},\ }\bibfield  {title}
  {\bibinfo {title} {Symmetry restoration and quantum mpemba effect in
  symmetric random circuits},\ }\href
  {https://doi.org/10.1103/PhysRevLett.133.140405} {\bibfield  {journal}
  {\bibinfo  {journal} {Phys. Rev. Lett.}\ }\textbf {\bibinfo {volume} {133}},\
  \bibinfo {pages} {140405} (\bibinfo {year} {2024}{\natexlab{a}})}\BibitemShut
  {NoStop}%
\bibitem [{\citenamefont {Ares}\ \emph
  {et~al.}(2025{\natexlab{b}})\citenamefont {Ares}, \citenamefont {Murciano},
  \citenamefont {Calabrese},\ and\ \citenamefont {Piroli}}]{QME9}%
  \BibitemOpen
  \bibfield  {author} {\bibinfo {author} {\bibfnamefont {F.}~\bibnamefont
  {Ares}}, \bibinfo {author} {\bibfnamefont {S.}~\bibnamefont {Murciano}},
  \bibinfo {author} {\bibfnamefont {P.}~\bibnamefont {Calabrese}},\ and\
  \bibinfo {author} {\bibfnamefont {L.}~\bibnamefont {Piroli}},\ }\bibfield
  {title} {\bibinfo {title} {Entanglement asymmetry dynamics in random quantum
  circuits},\ }\href {https://doi.org/10.1103/m3np-p5xj} {\bibfield  {journal}
  {\bibinfo  {journal} {Phys. Rev. Research}\ }\textbf {\bibinfo {volume}
  {7}},\ \bibinfo {pages} {033135} (\bibinfo {year}
  {2025}{\natexlab{b}})}\BibitemShut {NoStop}%
\bibitem [{\citenamefont {Turkeshi}\ \emph {et~al.}(2025)\citenamefont
  {Turkeshi}, \citenamefont {Calabrese},\ and\ \citenamefont
  {De~Luca}}]{QME901}%
  \BibitemOpen
  \bibfield  {author} {\bibinfo {author} {\bibfnamefont {X.}~\bibnamefont
  {Turkeshi}}, \bibinfo {author} {\bibfnamefont {P.}~\bibnamefont
  {Calabrese}},\ and\ \bibinfo {author} {\bibfnamefont {A.}~\bibnamefont
  {De~Luca}},\ }\bibfield  {title} {\bibinfo {title} {Quantum mpemba effect in
  random circuits},\ }\href {https://doi.org/10.1103/5d6p-8d1b} {\bibfield
  {journal} {\bibinfo  {journal} {Phys. Rev. Lett.}\ }\textbf {\bibinfo
  {volume} {135}},\ \bibinfo {pages} {040403} (\bibinfo {year}
  {2025})}\BibitemShut {NoStop}%
\bibitem [{\citenamefont {Klobas}\ \emph {et~al.}(2025)\citenamefont {Klobas},
  \citenamefont {Rylands},\ and\ \citenamefont {Bertini}}]{QME902}%
  \BibitemOpen
  \bibfield  {author} {\bibinfo {author} {\bibfnamefont {K.}~\bibnamefont
  {Klobas}}, \bibinfo {author} {\bibfnamefont {C.}~\bibnamefont {Rylands}},\
  and\ \bibinfo {author} {\bibfnamefont {B.}~\bibnamefont {Bertini}},\
  }\bibfield  {title} {\bibinfo {title} {Translation symmetry restoration under
  random unitary dynamics},\ }\href
  {https://doi.org/10.1103/PhysRevB.111.L140304} {\bibfield  {journal}
  {\bibinfo  {journal} {Phys. Rev. B}\ }\textbf {\bibinfo {volume} {111}},\
  \bibinfo {pages} {L140304} (\bibinfo {year} {2025})}\BibitemShut {NoStop}%
\bibitem [{\citenamefont {Klobas}(2024)}]{QME903}%
  \BibitemOpen
  \bibfield  {author} {\bibinfo {author} {\bibfnamefont {K.}~\bibnamefont
  {Klobas}},\ }\bibfield  {title} {\bibinfo {title} {Non-equilibrium dynamics
  of symmetry-resolved entanglement and entanglement asymmetry: exact
  asymptotics in rule 54},\ }\href {https://doi.org/10.1088/1751-8121/ad91fd}
  {\bibfield  {journal} {\bibinfo  {journal} {J. Phys. A: Math. Theor.}\
  }\textbf {\bibinfo {volume} {57}},\ \bibinfo {pages} {505001} (\bibinfo
  {year} {2024})}\BibitemShut {NoStop}%
\bibitem [{\citenamefont {Foligno}\ \emph {et~al.}(2025)\citenamefont
  {Foligno}, \citenamefont {Calabrese},\ and\ \citenamefont
  {Bertini}}]{QME904}%
  \BibitemOpen
  \bibfield  {author} {\bibinfo {author} {\bibfnamefont {A.}~\bibnamefont
  {Foligno}}, \bibinfo {author} {\bibfnamefont {P.}~\bibnamefont {Calabrese}},\
  and\ \bibinfo {author} {\bibfnamefont {B.}~\bibnamefont {Bertini}},\
  }\bibfield  {title} {\bibinfo {title} {Nonequilibrium dynamics of charged
  dual-unitary circuits},\ }\href {https://doi.org/10.1103/PRXQuantum.6.010324}
  {\bibfield  {journal} {\bibinfo  {journal} {PRX Quantum}\ }\textbf {\bibinfo
  {volume} {6}},\ \bibinfo {pages} {010324} (\bibinfo {year}
  {2025})}\BibitemShut {NoStop}%
\bibitem [{\citenamefont {Chatterjee}\ \emph {et~al.}(2023)\citenamefont
  {Chatterjee}, \citenamefont {Takada},\ and\ \citenamefont
  {Hayakawa}}]{QME_dot1}%
  \BibitemOpen
  \bibfield  {author} {\bibinfo {author} {\bibfnamefont {A.~K.}\ \bibnamefont
  {Chatterjee}}, \bibinfo {author} {\bibfnamefont {S.}~\bibnamefont {Takada}},\
  and\ \bibinfo {author} {\bibfnamefont {H.}~\bibnamefont {Hayakawa}},\
  }\bibfield  {title} {\bibinfo {title} {Quantum mpemba effect in a quantum dot
  with reservoirs},\ }\href {https://doi.org/10.1103/PhysRevLett.131.080402}
  {\bibfield  {journal} {\bibinfo  {journal} {Phys. Rev. Lett.}\ }\textbf
  {\bibinfo {volume} {131}},\ \bibinfo {pages} {080402} (\bibinfo {year}
  {2023})}\BibitemShut {NoStop}%
\bibitem [{\citenamefont {Graf}\ \emph {et~al.}(2025)\citenamefont {Graf},
  \citenamefont {Splettstoesser},\ and\ \citenamefont {Monsel}}]{QME_dot2}%
  \BibitemOpen
  \bibfield  {author} {\bibinfo {author} {\bibfnamefont {J.}~\bibnamefont
  {Graf}}, \bibinfo {author} {\bibfnamefont {J.}~\bibnamefont
  {Splettstoesser}},\ and\ \bibinfo {author} {\bibfnamefont {J.}~\bibnamefont
  {Monsel}},\ }\bibfield  {title} {\bibinfo {title} {Role of
  electron–electron interaction in the mpemba effect in quantum dots},\
  }\href {https://doi.org/10.1088/1361-648X/adc3e3} {\bibfield  {journal}
  {\bibinfo  {journal} {Journal of Physics: Condensed Matter}\ }\textbf
  {\bibinfo {volume} {37}},\ \bibinfo {pages} {195302} (\bibinfo {year}
  {2025})}\BibitemShut {NoStop}%
\bibitem [{\citenamefont {Carollo}\ \emph {et~al.}(2021)\citenamefont
  {Carollo}, \citenamefont {Lasanta},\ and\ \citenamefont
  {Lesanovsky}}]{OpenQME1}%
  \BibitemOpen
  \bibfield  {author} {\bibinfo {author} {\bibfnamefont {F.}~\bibnamefont
  {Carollo}}, \bibinfo {author} {\bibfnamefont {A.}~\bibnamefont {Lasanta}},\
  and\ \bibinfo {author} {\bibfnamefont {I.}~\bibnamefont {Lesanovsky}},\
  }\bibfield  {title} {\bibinfo {title} {Exponentially accelerated approach to
  stationarity in markovian open quantum systems through the mpemba effect},\
  }\href {https://doi.org/10.1103/PhysRevLett.127.060401} {\bibfield  {journal}
  {\bibinfo  {journal} {Phys. Rev. Lett.}\ }\textbf {\bibinfo {volume} {127}},\
  \bibinfo {pages} {060401} (\bibinfo {year} {2021})}\BibitemShut {NoStop}%
\bibitem [{\citenamefont {Kochsiek}\ \emph {et~al.}(2022)\citenamefont
  {Kochsiek}, \citenamefont {Carollo},\ and\ \citenamefont
  {Lesanovsky}}]{OpenQME2}%
  \BibitemOpen
  \bibfield  {author} {\bibinfo {author} {\bibfnamefont {S.}~\bibnamefont
  {Kochsiek}}, \bibinfo {author} {\bibfnamefont {F.}~\bibnamefont {Carollo}},\
  and\ \bibinfo {author} {\bibfnamefont {I.}~\bibnamefont {Lesanovsky}},\
  }\bibfield  {title} {\bibinfo {title} {Accelerating the approach of
  dissipative quantum spin systems towards stationarity through global spin
  rotations},\ }\href {https://doi.org/10.1103/PhysRevA.106.012207} {\bibfield
  {journal} {\bibinfo  {journal} {Phys. Rev. A}\ }\textbf {\bibinfo {volume}
  {106}},\ \bibinfo {pages} {012207} (\bibinfo {year} {2022})}\BibitemShut
  {NoStop}%
\bibitem [{\citenamefont {Liu}\ \emph {et~al.}(2024{\natexlab{b}})\citenamefont
  {Liu}, \citenamefont {Yuan}, \citenamefont {Ruan}, \citenamefont {Xu},
  \citenamefont {Luo}, \citenamefont {He}, \citenamefont {He}, \citenamefont
  {Ma},\ and\ \citenamefont {Wang}}]{OpenQME3}%
  \BibitemOpen
  \bibfield  {author} {\bibinfo {author} {\bibfnamefont {D.}~\bibnamefont
  {Liu}}, \bibinfo {author} {\bibfnamefont {J.}~\bibnamefont {Yuan}}, \bibinfo
  {author} {\bibfnamefont {H.}~\bibnamefont {Ruan}}, \bibinfo {author}
  {\bibfnamefont {Y.}~\bibnamefont {Xu}}, \bibinfo {author} {\bibfnamefont
  {S.}~\bibnamefont {Luo}}, \bibinfo {author} {\bibfnamefont {J.}~\bibnamefont
  {He}}, \bibinfo {author} {\bibfnamefont {X.}~\bibnamefont {He}}, \bibinfo
  {author} {\bibfnamefont {Y.}~\bibnamefont {Ma}},\ and\ \bibinfo {author}
  {\bibfnamefont {J.}~\bibnamefont {Wang}},\ }\bibfield  {title} {\bibinfo
  {title} {Speeding up quantum heat engines by the mpemba effect},\ }\href
  {https://doi.org/10.1103/PhysRevA.110.042218} {\bibfield  {journal} {\bibinfo
   {journal} {Phys. Rev. A}\ }\textbf {\bibinfo {volume} {110}},\ \bibinfo
  {pages} {042218} (\bibinfo {year} {2024}{\natexlab{b}})}\BibitemShut
  {NoStop}%
\bibitem [{\citenamefont {Furtado}\ and\ \citenamefont
  {Santos}(2025)}]{OpenQME4}%
  \BibitemOpen
  \bibfield  {author} {\bibinfo {author} {\bibfnamefont {J.}~\bibnamefont
  {Furtado}}\ and\ \bibinfo {author} {\bibfnamefont {A.~C.}\ \bibnamefont
  {Santos}},\ }\bibfield  {title} {\bibinfo {title} {Enhanced quantum mpemba
  effect with squeezed thermal reservoirs},\ }\href
  {https://doi.org/10.1016/j.aop.2025.170135} {\bibfield  {journal} {\bibinfo
  {journal} {Annals of Physics}\ }\textbf {\bibinfo {volume} {480}},\ \bibinfo
  {pages} {170135} (\bibinfo {year} {2025})}\BibitemShut {NoStop}%
\bibitem [{\citenamefont {Bettmann}\ and\ \citenamefont
  {Goold}(2025)}]{OpenQME5}%
  \BibitemOpen
  \bibfield  {author} {\bibinfo {author} {\bibfnamefont {L.~P.}\ \bibnamefont
  {Bettmann}}\ and\ \bibinfo {author} {\bibfnamefont {J.}~\bibnamefont
  {Goold}},\ }\bibfield  {title} {\bibinfo {title} {Information geometry
  approach to quantum stochastic thermodynamics},\ }\href
  {https://doi.org/10.1103/PhysRevE.111.014133} {\bibfield  {journal} {\bibinfo
   {journal} {Phys. Rev. E}\ }\textbf {\bibinfo {volume} {111}},\ \bibinfo
  {pages} {014133} (\bibinfo {year} {2025})}\BibitemShut {NoStop}%
\bibitem [{\citenamefont {Strachan}\ \emph {et~al.}(2025)\citenamefont
  {Strachan}, \citenamefont {Purkayastha},\ and\ \citenamefont
  {Clark}}]{OpenQME6}%
  \BibitemOpen
  \bibfield  {author} {\bibinfo {author} {\bibfnamefont {D.~J.}\ \bibnamefont
  {Strachan}}, \bibinfo {author} {\bibfnamefont {A.}~\bibnamefont
  {Purkayastha}},\ and\ \bibinfo {author} {\bibfnamefont {S.~R.}\ \bibnamefont
  {Clark}},\ }\bibfield  {title} {\bibinfo {title} {Non-markovian quantum
  mpemba effect},\ }\href {https://doi.org/10.1103/PhysRevLett.134.220403}
  {\bibfield  {journal} {\bibinfo  {journal} {Phys. Rev. Lett.}\ }\textbf
  {\bibinfo {volume} {134}},\ \bibinfo {pages} {220403} (\bibinfo {year}
  {2025})}\BibitemShut {NoStop}%
\bibitem [{\citenamefont {Medina}\ \emph {et~al.}(2025)\citenamefont {Medina},
  \citenamefont {Culhane}, \citenamefont {Binder}, \citenamefont {Landi},\ and\
  \citenamefont {Goold}}]{OpenQME7}%
  \BibitemOpen
  \bibfield  {author} {\bibinfo {author} {\bibfnamefont {I.}~\bibnamefont
  {Medina}}, \bibinfo {author} {\bibfnamefont {O.}~\bibnamefont {Culhane}},
  \bibinfo {author} {\bibfnamefont {F.~C.}\ \bibnamefont {Binder}}, \bibinfo
  {author} {\bibfnamefont {G.~T.}\ \bibnamefont {Landi}},\ and\ \bibinfo
  {author} {\bibfnamefont {J.}~\bibnamefont {Goold}},\ }\bibfield  {title}
  {\bibinfo {title} {Anomalous discharging of quantum batteries: The ergotropic
  mpemba effect},\ }\href {https://doi.org/10.1103/PhysRevLett.134.220402}
  {\bibfield  {journal} {\bibinfo  {journal} {Phys. Rev. Lett.}\ }\textbf
  {\bibinfo {volume} {134}},\ \bibinfo {pages} {220402} (\bibinfo {year}
  {2025})}\BibitemShut {NoStop}%
\bibitem [{\citenamefont {Wang}\ \emph
  {et~al.}(2024{\natexlab{a}})\citenamefont {Wang}, \citenamefont {Wu},
  \citenamefont {Byrd},\ and\ \citenamefont {Wu}}]{OpenQME8}%
  \BibitemOpen
  \bibfield  {author} {\bibinfo {author} {\bibfnamefont {Z.-M.}\ \bibnamefont
  {Wang}}, \bibinfo {author} {\bibfnamefont {S.~L.}\ \bibnamefont {Wu}},
  \bibinfo {author} {\bibfnamefont {M.~S.}\ \bibnamefont {Byrd}},\ and\
  \bibinfo {author} {\bibfnamefont {L.-A.}\ \bibnamefont {Wu}},\ }\bibfield
  {title} {\bibinfo {title} {Going beyond quantum markovianity and back to
  reality: An exact master equation study},\ }\href@noop {} {\bibfield
  {journal} {\bibinfo  {journal} {arXiv preprint}\ } (\bibinfo {year}
  {2024}{\natexlab{a}})},\ \Eprint {https://arxiv.org/abs/2411.17197}
  {arXiv:2411.17197} \BibitemShut {NoStop}%
\bibitem [{\citenamefont {Ma}\ and\ \citenamefont {Liu}(2025)}]{OpenQME10}%
  \BibitemOpen
  \bibfield  {author} {\bibinfo {author} {\bibfnamefont {W.}~\bibnamefont
  {Ma}}\ and\ \bibinfo {author} {\bibfnamefont {J.}~\bibnamefont {Liu}},\
  }\bibfield  {title} {\bibinfo {title} {Quantum mpemba effect in parity-time
  symmetric systems},\ }\href@noop {} {\bibfield  {journal} {\bibinfo
  {journal} {arXiv preprint}\ } (\bibinfo {year} {2025})},\ \Eprint
  {https://arxiv.org/abs/2508.17575} {arXiv:2508.17575} \BibitemShut {NoStop}%
\bibitem [{\citenamefont {Ali}\ \emph {et~al.}(2025)\citenamefont {Ali},
  \citenamefont {Hussain}, \citenamefont {Zad}, \citenamefont {Kuniyil},
  \citenamefont {Rahim}, \citenamefont {Al-Kuwari},\ and\ \citenamefont
  {Haddadi}}]{OpenQME11}%
  \BibitemOpen
  \bibfield  {author} {\bibinfo {author} {\bibfnamefont {A.}~\bibnamefont
  {Ali}}, \bibinfo {author} {\bibfnamefont {M.~I.}\ \bibnamefont {Hussain}},
  \bibinfo {author} {\bibfnamefont {H.~A.}\ \bibnamefont {Zad}}, \bibinfo
  {author} {\bibfnamefont {H.}~\bibnamefont {Kuniyil}}, \bibinfo {author}
  {\bibfnamefont {M.~T.}\ \bibnamefont {Rahim}}, \bibinfo {author}
  {\bibfnamefont {S.}~\bibnamefont {Al-Kuwari}},\ and\ \bibinfo {author}
  {\bibfnamefont {S.}~\bibnamefont {Haddadi}},\ }\bibfield  {title} {\bibinfo
  {title} {Quantum mpemba effect in a four-site bose-hubbard model},\
  }\href@noop {} {\bibfield  {journal} {\bibinfo  {journal} {arXiv preprint}\ }
  (\bibinfo {year} {2025})},\ \Eprint {https://arxiv.org/abs/2509.06937}
  {arXiv:2509.06937} \BibitemShut {NoStop}%
\bibitem [{\citenamefont {Li}\ \emph {et~al.}(2025)\citenamefont {Li},
  \citenamefont {Li},\ and\ \citenamefont {Li}}]{OpenQME12}%
  \BibitemOpen
  \bibfield  {author} {\bibinfo {author} {\bibfnamefont {Y.}~\bibnamefont
  {Li}}, \bibinfo {author} {\bibfnamefont {W.}~\bibnamefont {Li}},\ and\
  \bibinfo {author} {\bibfnamefont {X.}~\bibnamefont {Li}},\ }\bibfield
  {title} {\bibinfo {title} {Ergotropic mpemba effect in non-markovian quantum
  systems},\ }\href {https://doi.org/10.1103/5xrr-x2rm} {\bibfield  {journal}
  {\bibinfo  {journal} {Phys. Rev. A}\ }\textbf {\bibinfo {volume} {112}},\
  \bibinfo {pages} {032209} (\bibinfo {year} {2025})}\BibitemShut {NoStop}%
\bibitem [{\citenamefont {Chatterjee}\ \emph {et~al.}(2025)\citenamefont
  {Chatterjee}, \citenamefont {Khan}, \citenamefont {Jain},\ and\ \citenamefont
  {Mahesh}}]{OpenQME13}%
  \BibitemOpen
  \bibfield  {author} {\bibinfo {author} {\bibfnamefont {A.}~\bibnamefont
  {Chatterjee}}, \bibinfo {author} {\bibfnamefont {S.}~\bibnamefont {Khan}},
  \bibinfo {author} {\bibfnamefont {S.}~\bibnamefont {Jain}},\ and\ \bibinfo
  {author} {\bibfnamefont {T.~S.}\ \bibnamefont {Mahesh}},\ }\bibfield  {title}
  {\bibinfo {title} {Direct experimental observation of quantum mpemba effect
  without bath engineering},\ }\href@noop {} {\bibfield  {journal} {\bibinfo
  {journal} {arXiv preprint}\ } (\bibinfo {year} {2025})},\ \Eprint
  {https://arxiv.org/abs/2509.13451} {arXiv:2509.13451} \BibitemShut {NoStop}%
\bibitem [{\citenamefont {Wei}\ \emph {et~al.}(2026)\citenamefont {Wei},
  \citenamefont {Xu}, \citenamefont {Jiang}, \citenamefont {Hu},\ and\
  \citenamefont {Pan}}]{wei2025quantum}%
  \BibitemOpen
  \bibfield  {author} {\bibinfo {author} {\bibfnamefont {Z.}~\bibnamefont
  {Wei}}, \bibinfo {author} {\bibfnamefont {M.}~\bibnamefont {Xu}}, \bibinfo
  {author} {\bibfnamefont {X.-P.}\ \bibnamefont {Jiang}}, \bibinfo {author}
  {\bibfnamefont {H.}~\bibnamefont {Hu}},\ and\ \bibinfo {author}
  {\bibfnamefont {L.}~\bibnamefont {Pan}},\ }\bibfield  {title} {\bibinfo
  {title} {Quantum mpemba effect in dissipative spin chains at criticality},\
  }\href@noop {} {\bibfield  {journal} {\bibinfo  {journal} {Science China
  Physics, Mechanics \& Astronomy}\ }\textbf {\bibinfo {volume} {69}},\
  \bibinfo {pages} {240315} (\bibinfo {year} {2026})}\BibitemShut {NoStop}%
\bibitem [{\citenamefont {Caldas}\ and\ \citenamefont
  {Pires}(2025)}]{caldas2025exponentially}%
  \BibitemOpen
  \bibfield  {author} {\bibinfo {author} {\bibfnamefont {E.~L.}\ \bibnamefont
  {Caldas}}\ and\ \bibinfo {author} {\bibfnamefont {D.~P.}\ \bibnamefont
  {Pires}},\ }\bibfield  {title} {\bibinfo {title} {Exponentially accelerated
  relaxation and quantum mpemba effect in open quantum systems},\ }\href@noop
  {} {\bibfield  {journal} {\bibinfo  {journal} {arXiv preprint}\ } (\bibinfo
  {year} {2025})},\ \Eprint {https://arxiv.org/abs/2512.07561}
  {arXiv:2512.07561} \BibitemShut {NoStop}%
\bibitem [{\citenamefont {Liu}\ and\ \citenamefont
  {Wang}(2025)}]{liu2025general}%
  \BibitemOpen
  \bibfield  {author} {\bibinfo {author} {\bibfnamefont {Y.}~\bibnamefont
  {Liu}}\ and\ \bibinfo {author} {\bibfnamefont {Y.}~\bibnamefont {Wang}},\
  }\bibfield  {title} {\bibinfo {title} {A general strategy for realizing
  mpemba effects in open quantum systems},\ }\href@noop {} {\bibfield
  {journal} {\bibinfo  {journal} {arXiv preprint}\ } (\bibinfo {year}
  {2025})},\ \Eprint {https://arxiv.org/abs/2511.04354} {arXiv:2511.04354}
  \BibitemShut {NoStop}%
\bibitem [{\citenamefont {Fossati}\ \emph {et~al.}(2024)\citenamefont
  {Fossati}, \citenamefont {Rylands},\ and\ \citenamefont {Calabrese}}]{QME10}%
  \BibitemOpen
  \bibfield  {author} {\bibinfo {author} {\bibfnamefont {M.}~\bibnamefont
  {Fossati}}, \bibinfo {author} {\bibfnamefont {C.}~\bibnamefont {Rylands}},\
  and\ \bibinfo {author} {\bibfnamefont {P.}~\bibnamefont {Calabrese}},\
  }\bibfield  {title} {\bibinfo {title} {Entanglement asymmetry in cft with
  boundary symmetry breaking},\ }\href@noop {} {\bibfield  {journal} {\bibinfo
  {journal} {arXiv preprint}\ } (\bibinfo {year} {2024})},\ \Eprint
  {https://arxiv.org/abs/2411.10244} {arXiv:2411.10244} \BibitemShut {NoStop}%
\bibitem [{\citenamefont {Chang}\ \emph {et~al.}(2024)\citenamefont {Chang},
  \citenamefont {Yin}, \citenamefont {Zhang},\ and\ \citenamefont
  {Li}}]{QME105}%
  \BibitemOpen
  \bibfield  {author} {\bibinfo {author} {\bibfnamefont {W.-X.}\ \bibnamefont
  {Chang}}, \bibinfo {author} {\bibfnamefont {S.}~\bibnamefont {Yin}}, \bibinfo
  {author} {\bibfnamefont {S.-X.}\ \bibnamefont {Zhang}},\ and\ \bibinfo
  {author} {\bibfnamefont {Z.-X.}\ \bibnamefont {Li}},\ }\bibfield  {title}
  {\bibinfo {title} {Imaginary-time mpemba effect in quantum many-body
  system},\ }\href@noop {} {\bibfield  {journal} {\bibinfo  {journal} {arXiv
  preprint}\ } (\bibinfo {year} {2024})},\ \Eprint
  {https://arxiv.org/abs/2409.06547} {arXiv:2409.06547} \BibitemShut {NoStop}%
\bibitem [{\citenamefont {Longhi}(2025)}]{QME12}%
  \BibitemOpen
  \bibfield  {author} {\bibinfo {author} {\bibfnamefont {S.}~\bibnamefont
  {Longhi}},\ }\bibfield  {title} {\bibinfo {title} {Mpemba effect and
  super-accelerated thermalization in the damped quantum harmonic oscillator},\
  }\href {https://doi.org/10.22331/q-2025-03-26-1677} {\bibfield  {journal}
  {\bibinfo  {journal} {Quantum}\ }\textbf {\bibinfo {volume} {9}},\ \bibinfo
  {pages} {1677} (\bibinfo {year} {2025})}\BibitemShut {NoStop}%
\bibitem [{\citenamefont {Wang}\ \emph
  {et~al.}(2024{\natexlab{b}})\citenamefont {Wang}, \citenamefont {Su},\ and\
  \citenamefont {Wang}}]{QME15}%
  \BibitemOpen
  \bibfield  {author} {\bibinfo {author} {\bibfnamefont {X.}~\bibnamefont
  {Wang}}, \bibinfo {author} {\bibfnamefont {J.}~\bibnamefont {Su}},\ and\
  \bibinfo {author} {\bibfnamefont {J.}~\bibnamefont {Wang}},\ }\bibfield
  {title} {\bibinfo {title} {Mpemba meets quantum chaos: Anomalous relaxation
  and mpemba crossings in dissipative sachdev-ye-kitaev models},\ }\href@noop
  {} {\bibfield  {journal} {\bibinfo  {journal} {arXiv preprint}\ } (\bibinfo
  {year} {2024}{\natexlab{b}})},\ \Eprint {https://arxiv.org/abs/2410.06669}
  {arXiv:2410.06669} \BibitemShut {NoStop}%
\bibitem [{\citenamefont {Vu}\ and\ \citenamefont {Hayakawa}(2025)}]{QME20}%
  \BibitemOpen
  \bibfield  {author} {\bibinfo {author} {\bibfnamefont {T.~V.}\ \bibnamefont
  {Vu}}\ and\ \bibinfo {author} {\bibfnamefont {H.}~\bibnamefont {Hayakawa}},\
  }\bibfield  {title} {\bibinfo {title} {Thermomajorization mpemba effect},\
  }\href {https://doi.org/10.1103/PhysRevLett.134.107101} {\bibfield  {journal}
  {\bibinfo  {journal} {Phys. Rev. Lett.}\ }\textbf {\bibinfo {volume} {134}},\
  \bibinfo {pages} {107101} (\bibinfo {year} {2025})}\BibitemShut {NoStop}%
\bibitem [{\citenamefont {Summer}\ \emph {et~al.}(2025)\citenamefont {Summer},
  \citenamefont {Moroder}, \citenamefont {Bettmann}, \citenamefont {Turkeshi},
  \citenamefont {Marvian},\ and\ \citenamefont {Goold}}]{QME21}%
  \BibitemOpen
  \bibfield  {author} {\bibinfo {author} {\bibfnamefont {A.}~\bibnamefont
  {Summer}}, \bibinfo {author} {\bibfnamefont {M.}~\bibnamefont {Moroder}},
  \bibinfo {author} {\bibfnamefont {L.~P.}\ \bibnamefont {Bettmann}}, \bibinfo
  {author} {\bibfnamefont {X.}~\bibnamefont {Turkeshi}}, \bibinfo {author}
  {\bibfnamefont {I.}~\bibnamefont {Marvian}},\ and\ \bibinfo {author}
  {\bibfnamefont {J.}~\bibnamefont {Goold}},\ }\bibfield  {title} {\bibinfo
  {title} {A resource theoretical unification of mpemba effects: classical and
  quantum},\ }\href@noop {} {\bibfield  {journal} {\bibinfo  {journal} {arXiv
  preprint}\ } (\bibinfo {year} {2025})},\ \Eprint
  {https://arxiv.org/abs/2507.16976} {arXiv:2507.16976} \BibitemShut {NoStop}%
\bibitem [{\citenamefont {Bagui}\ \emph {et~al.}(2025)\citenamefont {Bagui},
  \citenamefont {Chatterjee},\ and\ \citenamefont {Agarwalla}}]{QME26}%
  \BibitemOpen
  \bibfield  {author} {\bibinfo {author} {\bibfnamefont {P.}~\bibnamefont
  {Bagui}}, \bibinfo {author} {\bibfnamefont {A.}~\bibnamefont {Chatterjee}},\
  and\ \bibinfo {author} {\bibfnamefont {B.~K.}\ \bibnamefont {Agarwalla}},\
  }\bibfield  {title} {\bibinfo {title} {Accelerated relaxation and mpemba-like
  effect for operators in open quantum systems},\ }\href@noop {} {\bibfield
  {journal} {\bibinfo  {journal} {arXiv preprint}\ } (\bibinfo {year}
  {2025})},\ \Eprint {https://arxiv.org/abs/2510.24630} {arXiv:2510.24630}
  \BibitemShut {NoStop}%
\bibitem [{\citenamefont {Bao}\ and\ \citenamefont {Hou}(2025)}]{QME30}%
  \BibitemOpen
  \bibfield  {author} {\bibinfo {author} {\bibfnamefont {R.}~\bibnamefont
  {Bao}}\ and\ \bibinfo {author} {\bibfnamefont {Z.}~\bibnamefont {Hou}},\
  }\bibfield  {title} {\bibinfo {title} {Accelerating quantum relaxation via
  temporary reset: A mpemba-inspired approach},\ }\href
  {https://doi.org/10.1103/g94p-7421} {\bibfield  {journal} {\bibinfo
  {journal} {Phys. Rev. Lett.}\ }\textbf {\bibinfo {volume} {135}},\ \bibinfo
  {pages} {150403} (\bibinfo {year} {2025})}\BibitemShut {NoStop}%
\bibitem [{\citenamefont {Hallam}\ \emph {et~al.}(2025)\citenamefont {Hallam},
  \citenamefont {Yusuf}, \citenamefont {Clerk}, \citenamefont {Martin},\ and\
  \citenamefont {Papić}}]{QME31}%
  \BibitemOpen
  \bibfield  {author} {\bibinfo {author} {\bibfnamefont {A.}~\bibnamefont
  {Hallam}}, \bibinfo {author} {\bibfnamefont {M.}~\bibnamefont {Yusuf}},
  \bibinfo {author} {\bibfnamefont {A.~A.}\ \bibnamefont {Clerk}}, \bibinfo
  {author} {\bibfnamefont {I.}~\bibnamefont {Martin}},\ and\ \bibinfo {author}
  {\bibfnamefont {Z.}~\bibnamefont {Papić}},\ }\bibfield  {title} {\bibinfo
  {title} {Tunable quantum mpemba effect in long-range interacting systems},\
  }\href@noop {} {\bibfield  {journal} {\bibinfo  {journal} {arXiv preprint}\ }
  (\bibinfo {year} {2025})},\ \Eprint {https://arxiv.org/abs/2510.12875}
  {arXiv:2510.12875} \BibitemShut {NoStop}%
\bibitem [{\citenamefont {Kheirandish}\ \emph {et~al.}(2025)\citenamefont
  {Kheirandish}, \citenamefont {Cheraghpour},\ and\ \citenamefont
  {Moradian}}]{R1}%
  \BibitemOpen
  \bibfield  {author} {\bibinfo {author} {\bibfnamefont {F.}~\bibnamefont
  {Kheirandish}}, \bibinfo {author} {\bibfnamefont {N.}~\bibnamefont
  {Cheraghpour}},\ and\ \bibinfo {author} {\bibfnamefont {A.}~\bibnamefont
  {Moradian}},\ }\bibfield  {title} {\bibinfo {title} {The mpemba effect in
  quantum oscillating and two-level systems},\ }\href
  {https://doi.org/10.1016/j.physleta.2025.130915} {\bibfield  {journal}
  {\bibinfo  {journal} {Phys. Lett. A}\ }\textbf {\bibinfo {volume} {559}},\
  \bibinfo {pages} {130915} (\bibinfo {year} {2025})}\BibitemShut {NoStop}%
\bibitem [{\citenamefont {Nava}\ and\ \citenamefont {Egger}(2024)}]{R2}%
  \BibitemOpen
  \bibfield  {author} {\bibinfo {author} {\bibfnamefont {A.}~\bibnamefont
  {Nava}}\ and\ \bibinfo {author} {\bibfnamefont {R.}~\bibnamefont {Egger}},\
  }\bibfield  {title} {\bibinfo {title} {Mpemba effects in open nonequilibrium
  quantum systems},\ }\href {https://doi.org/10.1103/PhysRevLett.133.136302}
  {\bibfield  {journal} {\bibinfo  {journal} {Phys. Rev. Lett.}\ }\textbf
  {\bibinfo {volume} {133}},\ \bibinfo {pages} {136302} (\bibinfo {year}
  {2024})}\BibitemShut {NoStop}%
\bibitem [{\citenamefont {Barontini}\ \emph {et~al.}(2013)\citenamefont
  {Barontini}, \citenamefont {Labouvie}, \citenamefont {Stubenrauch},
  \citenamefont {Vogler}, \citenamefont {Guarrera},\ and\ \citenamefont
  {Ott}}]{Exp1}%
  \BibitemOpen
  \bibfield  {author} {\bibinfo {author} {\bibfnamefont {G.}~\bibnamefont
  {Barontini}}, \bibinfo {author} {\bibfnamefont {R.}~\bibnamefont {Labouvie}},
  \bibinfo {author} {\bibfnamefont {F.}~\bibnamefont {Stubenrauch}}, \bibinfo
  {author} {\bibfnamefont {A.}~\bibnamefont {Vogler}}, \bibinfo {author}
  {\bibfnamefont {V.}~\bibnamefont {Guarrera}},\ and\ \bibinfo {author}
  {\bibfnamefont {H.}~\bibnamefont {Ott}},\ }\bibfield  {title} {\bibinfo
  {title} {Controlling the dynamics of an open many-body quantum system with
  localized dissipation},\ }\href
  {https://doi.org/10.1103/PhysRevLett.110.035302} {\bibfield  {journal}
  {\bibinfo  {journal} {Phys. Rev. Lett.}\ }\textbf {\bibinfo {volume} {110}},\
  \bibinfo {pages} {035302} (\bibinfo {year} {2013})}\BibitemShut {NoStop}%
\bibitem [{\citenamefont {Yan}\ \emph {et~al.}(2013)\citenamefont {Yan},
  \citenamefont {Moses}, \citenamefont {Gadway}, \citenamefont {Covey},
  \citenamefont {Hazzard}, \citenamefont {Rey}, \citenamefont {Jin},\ and\
  \citenamefont {Ye}}]{Exp2}%
  \BibitemOpen
  \bibfield  {author} {\bibinfo {author} {\bibfnamefont {B.}~\bibnamefont
  {Yan}}, \bibinfo {author} {\bibfnamefont {S.~A.}\ \bibnamefont {Moses}},
  \bibinfo {author} {\bibfnamefont {B.}~\bibnamefont {Gadway}}, \bibinfo
  {author} {\bibfnamefont {J.~P.}\ \bibnamefont {Covey}}, \bibinfo {author}
  {\bibfnamefont {K.~R.}\ \bibnamefont {Hazzard}}, \bibinfo {author}
  {\bibfnamefont {A.~M.}\ \bibnamefont {Rey}}, \bibinfo {author} {\bibfnamefont
  {D.~S.}\ \bibnamefont {Jin}},\ and\ \bibinfo {author} {\bibfnamefont
  {J.}~\bibnamefont {Ye}},\ }\bibfield  {title} {\bibinfo {title} {Observation
  of dipolar spin-exchange interactions with lattice-confined polar
  molecules},\ }\href {https://doi.org/10.1038/nature12483} {\bibfield
  {journal} {\bibinfo  {journal} {Nature (London)}\ }\textbf {\bibinfo {volume}
  {501}},\ \bibinfo {pages} {521} (\bibinfo {year} {2013})}\BibitemShut
  {NoStop}%
\bibitem [{\citenamefont {Patil}\ \emph {et~al.}(2015)\citenamefont {Patil},
  \citenamefont {Chakram},\ and\ \citenamefont {Vengalattore}}]{Exp3}%
  \BibitemOpen
  \bibfield  {author} {\bibinfo {author} {\bibfnamefont {Y.~S.}\ \bibnamefont
  {Patil}}, \bibinfo {author} {\bibfnamefont {S.}~\bibnamefont {Chakram}},\
  and\ \bibinfo {author} {\bibfnamefont {M.}~\bibnamefont {Vengalattore}},\
  }\bibfield  {title} {\bibinfo {title} {Measurement-induced localization of an
  ultracold lattice gas},\ }\href
  {https://doi.org/10.1103/PhysRevLett.115.140402} {\bibfield  {journal}
  {\bibinfo  {journal} {Phys. Rev. Lett.}\ }\textbf {\bibinfo {volume} {115}},\
  \bibinfo {pages} {140402} (\bibinfo {year} {2015})}\BibitemShut {NoStop}%
\bibitem [{\citenamefont {Labouvie}\ \emph {et~al.}(2016)\citenamefont
  {Labouvie}, \citenamefont {Santra}, \citenamefont {Heun},\ and\ \citenamefont
  {Ott}}]{Exp4}%
  \BibitemOpen
  \bibfield  {author} {\bibinfo {author} {\bibfnamefont {R.}~\bibnamefont
  {Labouvie}}, \bibinfo {author} {\bibfnamefont {B.}~\bibnamefont {Santra}},
  \bibinfo {author} {\bibfnamefont {S.}~\bibnamefont {Heun}},\ and\ \bibinfo
  {author} {\bibfnamefont {H.}~\bibnamefont {Ott}},\ }\bibfield  {title}
  {\bibinfo {title} {Bistability in a driven-dissipative superfluid},\ }\href
  {https://doi.org/10.1103/PhysRevLett.116.235302} {\bibfield  {journal}
  {\bibinfo  {journal} {Phys. Rev. Lett.}\ }\textbf {\bibinfo {volume} {116}},\
  \bibinfo {pages} {235302} (\bibinfo {year} {2016})}\BibitemShut {NoStop}%
\bibitem [{\citenamefont {Tomita}\ \emph {et~al.}(2017)\citenamefont {Tomita},
  \citenamefont {Nakajima}, \citenamefont {Danshita}, \citenamefont {Takasu},\
  and\ \citenamefont {Takahashi}}]{Exp5}%
  \BibitemOpen
  \bibfield  {author} {\bibinfo {author} {\bibfnamefont {T.}~\bibnamefont
  {Tomita}}, \bibinfo {author} {\bibfnamefont {S.}~\bibnamefont {Nakajima}},
  \bibinfo {author} {\bibfnamefont {I.}~\bibnamefont {Danshita}}, \bibinfo
  {author} {\bibfnamefont {Y.}~\bibnamefont {Takasu}},\ and\ \bibinfo {author}
  {\bibfnamefont {Y.}~\bibnamefont {Takahashi}},\ }\bibfield  {title} {\bibinfo
  {title} {Observation of the mott insulator to superfluid crossover of a
  driven-dissipative bose-hubbard system},\ }\href
  {https://doi.org/10.1126/sciadv.1701513} {\bibfield  {journal} {\bibinfo
  {journal} {Sci. Adv.}\ }\textbf {\bibinfo {volume} {3}},\ \bibinfo {pages}
  {e1701513} (\bibinfo {year} {2017})}\BibitemShut {NoStop}%
\bibitem [{\citenamefont {Lüschen}\ \emph {et~al.}(2017)\citenamefont
  {Lüschen}, \citenamefont {Bordia}, \citenamefont {Hodgman}, \citenamefont
  {Schreiber}, \citenamefont {Sarkar}, \citenamefont {Daley}, \citenamefont
  {Fischer}, \citenamefont {Altman}, \citenamefont {Bloch},\ and\ \citenamefont
  {Schneider}}]{Exp6}%
  \BibitemOpen
  \bibfield  {author} {\bibinfo {author} {\bibfnamefont {H.~P.}\ \bibnamefont
  {Lüschen}}, \bibinfo {author} {\bibfnamefont {P.}~\bibnamefont {Bordia}},
  \bibinfo {author} {\bibfnamefont {S.~S.}\ \bibnamefont {Hodgman}}, \bibinfo
  {author} {\bibfnamefont {M.}~\bibnamefont {Schreiber}}, \bibinfo {author}
  {\bibfnamefont {S.}~\bibnamefont {Sarkar}}, \bibinfo {author} {\bibfnamefont
  {A.~J.}\ \bibnamefont {Daley}}, \bibinfo {author} {\bibfnamefont {M.~H.}\
  \bibnamefont {Fischer}}, \bibinfo {author} {\bibfnamefont {E.}~\bibnamefont
  {Altman}}, \bibinfo {author} {\bibfnamefont {I.}~\bibnamefont {Bloch}},\ and\
  \bibinfo {author} {\bibfnamefont {U.}~\bibnamefont {Schneider}},\ }\bibfield
  {title} {\bibinfo {title} {Signatures of many-body localization in a
  controlled open quantum system},\ }\href
  {https://doi.org/10.1103/PhysRevX.7.011034} {\bibfield  {journal} {\bibinfo
  {journal} {Phys. Rev. X}\ }\textbf {\bibinfo {volume} {7}},\ \bibinfo {pages}
  {011034} (\bibinfo {year} {2017})}\BibitemShut {NoStop}%
\bibitem [{\citenamefont {Sponselee}\ \emph {et~al.}(2018)\citenamefont
  {Sponselee}, \citenamefont {Freystatzky}, \citenamefont {Abeln},
  \citenamefont {Diem}, \citenamefont {Hundt}, \citenamefont {Kochanke},
  \citenamefont {Ponath}, \citenamefont {Santra}, \citenamefont {Mathey},
  \citenamefont {Sengstock},\ and\ \citenamefont {Becker}}]{Exp7}%
  \BibitemOpen
  \bibfield  {author} {\bibinfo {author} {\bibfnamefont {K.}~\bibnamefont
  {Sponselee}}, \bibinfo {author} {\bibfnamefont {L.}~\bibnamefont
  {Freystatzky}}, \bibinfo {author} {\bibfnamefont {B.}~\bibnamefont {Abeln}},
  \bibinfo {author} {\bibfnamefont {M.}~\bibnamefont {Diem}}, \bibinfo {author}
  {\bibfnamefont {B.}~\bibnamefont {Hundt}}, \bibinfo {author} {\bibfnamefont
  {A.}~\bibnamefont {Kochanke}}, \bibinfo {author} {\bibfnamefont
  {T.}~\bibnamefont {Ponath}}, \bibinfo {author} {\bibfnamefont
  {B.}~\bibnamefont {Santra}}, \bibinfo {author} {\bibfnamefont
  {L.}~\bibnamefont {Mathey}}, \bibinfo {author} {\bibfnamefont
  {K.}~\bibnamefont {Sengstock}},\ and\ \bibinfo {author} {\bibfnamefont
  {C.}~\bibnamefont {Becker}},\ }\bibfield  {title} {\bibinfo {title} {Dynamics
  of ultracold quantum gases in the dissipative fermi-hubbard model},\ }\href
  {https://doi.org/10.1088/2058-9565/aadccd} {\bibfield  {journal} {\bibinfo
  {journal} {Quantum Sci. Technol.}\ }\textbf {\bibinfo {volume} {4}},\
  \bibinfo {pages} {014002} (\bibinfo {year} {2018})}\BibitemShut {NoStop}%
\bibitem [{\citenamefont {Tomita}\ \emph {et~al.}(2019)\citenamefont {Tomita},
  \citenamefont {Nakajima}, \citenamefont {Takasu},\ and\ \citenamefont
  {Takahashi}}]{Exp8}%
  \BibitemOpen
  \bibfield  {author} {\bibinfo {author} {\bibfnamefont {T.}~\bibnamefont
  {Tomita}}, \bibinfo {author} {\bibfnamefont {S.}~\bibnamefont {Nakajima}},
  \bibinfo {author} {\bibfnamefont {Y.}~\bibnamefont {Takasu}},\ and\ \bibinfo
  {author} {\bibfnamefont {Y.}~\bibnamefont {Takahashi}},\ }\bibfield  {title}
  {\bibinfo {title} {Dissipative bose-hubbard system with intrinsic two-body
  loss},\ }\href {https://doi.org/10.1103/PhysRevA.99.031601} {\bibfield
  {journal} {\bibinfo  {journal} {Phys. Rev. A}\ }\textbf {\bibinfo {volume}
  {99}},\ \bibinfo {pages} {031601(R)} (\bibinfo {year} {2019})}\BibitemShut
  {NoStop}%
\bibitem [{\citenamefont {Takasu}\ \emph {et~al.}(2020)\citenamefont {Takasu},
  \citenamefont {Yagami}, \citenamefont {Ashida}, \citenamefont {Hamazaki},
  \citenamefont {Kuno},\ and\ \citenamefont {Takahashi}}]{Exp_new1}%
  \BibitemOpen
  \bibfield  {author} {\bibinfo {author} {\bibfnamefont {Y.}~\bibnamefont
  {Takasu}}, \bibinfo {author} {\bibfnamefont {T.}~\bibnamefont {Yagami}},
  \bibinfo {author} {\bibfnamefont {Y.}~\bibnamefont {Ashida}}, \bibinfo
  {author} {\bibfnamefont {R.}~\bibnamefont {Hamazaki}}, \bibinfo {author}
  {\bibfnamefont {Y.}~\bibnamefont {Kuno}},\ and\ \bibinfo {author}
  {\bibfnamefont {Y.}~\bibnamefont {Takahashi}},\ }\bibfield  {title} {\bibinfo
  {title} {Pt-symmetric non-hermitian quantum many-body system using ultracold
  atoms in an optical lattice with controlled dissipation},\ }\href
  {https://doi.org/10.1093/ptep/ptaa178} {\bibfield  {journal} {\bibinfo
  {journal} {Prog. Theor. Exp. Phys.}\ }\textbf {\bibinfo {volume} {2020}},\
  \bibinfo {pages} {12A110} (\bibinfo {year} {2020})}\BibitemShut {NoStop}%
\bibitem [{\citenamefont {Bouganne}\ \emph {et~al.}(2020)\citenamefont
  {Bouganne}, \citenamefont {Aguilera}, \citenamefont {Ghermaoui},\ and\
  \citenamefont {Gerbier}}]{Exp_new2}%
  \BibitemOpen
  \bibfield  {author} {\bibinfo {author} {\bibfnamefont {R.}~\bibnamefont
  {Bouganne}}, \bibinfo {author} {\bibfnamefont {M.~B.}\ \bibnamefont
  {Aguilera}}, \bibinfo {author} {\bibfnamefont {A.}~\bibnamefont
  {Ghermaoui}},\ and\ \bibinfo {author} {\bibfnamefont {F.}~\bibnamefont
  {Gerbier}},\ }\bibfield  {title} {\bibinfo {title} {Anomalous decay of
  coherence in a dissipative many-body system},\ }\href
  {https://doi.org/10.1038/s41567-019-0678-2} {\bibfield  {journal} {\bibinfo
  {journal} {Nat. Phys.}\ }\textbf {\bibinfo {volume} {16}},\ \bibinfo {pages}
  {21} (\bibinfo {year} {2020})}\BibitemShut {NoStop}%
\bibitem [{\citenamefont {Zhao}\ \emph {et~al.}(2023)\citenamefont {Zhao},
  \citenamefont {Tian}, \citenamefont {Ye}, \citenamefont {Wu}, \citenamefont
  {Zhao}, \citenamefont {Chi}, \citenamefont {Tian}, \citenamefont {Yao},
  \citenamefont {Hu}, \citenamefont {Chen},\ and\ \citenamefont
  {Chen}}]{Exp_new3}%
  \BibitemOpen
  \bibfield  {author} {\bibinfo {author} {\bibfnamefont {Y.}~\bibnamefont
  {Zhao}}, \bibinfo {author} {\bibfnamefont {Y.}~\bibnamefont {Tian}}, \bibinfo
  {author} {\bibfnamefont {J.}~\bibnamefont {Ye}}, \bibinfo {author}
  {\bibfnamefont {Y.}~\bibnamefont {Wu}}, \bibinfo {author} {\bibfnamefont
  {Z.}~\bibnamefont {Zhao}}, \bibinfo {author} {\bibfnamefont {Z.}~\bibnamefont
  {Chi}}, \bibinfo {author} {\bibfnamefont {T.}~\bibnamefont {Tian}}, \bibinfo
  {author} {\bibfnamefont {H.}~\bibnamefont {Yao}}, \bibinfo {author}
  {\bibfnamefont {J.}~\bibnamefont {Hu}}, \bibinfo {author} {\bibfnamefont
  {Y.}~\bibnamefont {Chen}},\ and\ \bibinfo {author} {\bibfnamefont
  {W.}~\bibnamefont {Chen}},\ }\bibfield  {title} {\bibinfo {title}
  {Observation of universal dissipative dynamics in strongly correlated quantum
  gas},\ }\href@noop {} {\bibfield  {journal} {\bibinfo  {journal} {arXiv
  preprint}\ } (\bibinfo {year} {2023})},\ \Eprint
  {https://arxiv.org/abs/2309.10257} {arXiv:2309.10257} \BibitemShut {NoStop}%
\bibitem [{\citenamefont {Nava}\ and\ \citenamefont {Egger}(2025)}]{QME27}%
  \BibitemOpen
  \bibfield  {author} {\bibinfo {author} {\bibfnamefont {A.}~\bibnamefont
  {Nava}}\ and\ \bibinfo {author} {\bibfnamefont {R.}~\bibnamefont {Egger}},\
  }\bibfield  {title} {\bibinfo {title} {Pontus-mpemba effects},\ }\href
  {https://doi.org/10.1103/hhgj-89gj} {\bibfield  {journal} {\bibinfo
  {journal} {Phys. Rev. Lett.}\ }\textbf {\bibinfo {volume} {135}},\ \bibinfo
  {pages} {140404} (\bibinfo {year} {2025})}\BibitemShut {NoStop}%
\bibitem [{\citenamefont {Longhi}(2026)}]{Longhi_2026}%
  \BibitemOpen
  \bibfield  {author} {\bibinfo {author} {\bibfnamefont {S.}~\bibnamefont
  {Longhi}},\ }\bibfield  {title} {\bibinfo {title} {Quantum pontus–mpemba
  effect enabled by the liouvillian skin effect},\ }\href
  {https://doi.org/10.1088/1751-8121/ae4079} {\bibfield  {journal} {\bibinfo
  {journal} {Journal of Physics A: Mathematical and Theoretical}\ }\textbf
  {\bibinfo {volume} {59}},\ \bibinfo {pages} {065304} (\bibinfo {year}
  {2026})}\BibitemShut {NoStop}%
\bibitem [{\citenamefont {Nava}\ \emph {et~al.}(2025)\citenamefont {Nava},
  \citenamefont {Egger}, \citenamefont {Dey},\ and\ \citenamefont
  {Giuliano}}]{QME28}%
  \BibitemOpen
  \bibfield  {author} {\bibinfo {author} {\bibfnamefont {A.}~\bibnamefont
  {Nava}}, \bibinfo {author} {\bibfnamefont {R.}~\bibnamefont {Egger}},
  \bibinfo {author} {\bibfnamefont {B.}~\bibnamefont {Dey}},\ and\ \bibinfo
  {author} {\bibfnamefont {D.}~\bibnamefont {Giuliano}},\ }\bibfield  {title}
  {\bibinfo {title} {Speeding up pontus-mpemba effects via dynamical phase
  transitions},\ }\href@noop {} {\bibfield  {journal} {\bibinfo  {journal}
  {arXiv preprint}\ } (\bibinfo {year} {2025})},\ \Eprint
  {https://arxiv.org/abs/2509.09366} {arXiv:2509.09366} \BibitemShut {NoStop}%
\bibitem [{\citenamefont {Yu}\ \emph {et~al.}(2025{\natexlab{c}})\citenamefont
  {Yu}, \citenamefont {Hu},\ and\ \citenamefont {Zhang}}]{QME29}%
  \BibitemOpen
  \bibfield  {author} {\bibinfo {author} {\bibfnamefont {H.}~\bibnamefont
  {Yu}}, \bibinfo {author} {\bibfnamefont {J.}~\bibnamefont {Hu}},\ and\
  \bibinfo {author} {\bibfnamefont {S.-X.}\ \bibnamefont {Zhang}},\ }\bibfield
  {title} {\bibinfo {title} {Quantum pontus-mpemba effects in real and
  imaginary-time dynamics},\ }\href@noop {} {\bibfield  {journal} {\bibinfo
  {journal} {arXiv preprint}\ } (\bibinfo {year} {2025}{\natexlab{c}})},\
  \Eprint {https://arxiv.org/abs/2509.01960} {arXiv:2509.01960} \BibitemShut
  {NoStop}%
\bibitem [{\citenamefont {Yusipov}\ \emph {et~al.}(2017)\citenamefont
  {Yusipov}, \citenamefont {Laptyeva}, \citenamefont {Denisov},\ and\
  \citenamefont {Ivanchenko}}]{Yusipov17}%
  \BibitemOpen
  \bibfield  {author} {\bibinfo {author} {\bibfnamefont {I.}~\bibnamefont
  {Yusipov}}, \bibinfo {author} {\bibfnamefont {T.}~\bibnamefont {Laptyeva}},
  \bibinfo {author} {\bibfnamefont {S.}~\bibnamefont {Denisov}},\ and\ \bibinfo
  {author} {\bibfnamefont {M.}~\bibnamefont {Ivanchenko}},\ }\bibfield  {title}
  {\bibinfo {title} {Localization in open quantum systems},\ }\href
  {https://doi.org/10.1103/PhysRevLett.118.070402} {\bibfield  {journal}
  {\bibinfo  {journal} {Phys. Rev. Lett.}\ }\textbf {\bibinfo {volume} {118}},\
  \bibinfo {pages} {070402} (\bibinfo {year} {2017})}\BibitemShut {NoStop}%
\bibitem [{\citenamefont {Vakulchyk}\ \emph {et~al.}(2018)\citenamefont
  {Vakulchyk}, \citenamefont {Yusipov}, \citenamefont {Ivanchenko},
  \citenamefont {Flach},\ and\ \citenamefont {Denisov}}]{Yusipov18}%
  \BibitemOpen
  \bibfield  {author} {\bibinfo {author} {\bibfnamefont {I.}~\bibnamefont
  {Vakulchyk}}, \bibinfo {author} {\bibfnamefont {I.}~\bibnamefont {Yusipov}},
  \bibinfo {author} {\bibfnamefont {M.}~\bibnamefont {Ivanchenko}}, \bibinfo
  {author} {\bibfnamefont {S.}~\bibnamefont {Flach}},\ and\ \bibinfo {author}
  {\bibfnamefont {S.}~\bibnamefont {Denisov}},\ }\bibfield  {title} {\bibinfo
  {title} {Signatures of many-body localization in steady states of open
  quantum systems},\ }\href {https://doi.org/10.1103/PhysRevB.98.020202}
  {\bibfield  {journal} {\bibinfo  {journal} {Phys. Rev. B}\ }\textbf {\bibinfo
  {volume} {98}},\ \bibinfo {pages} {020202} (\bibinfo {year} {2018})},\
  \bibinfo {note} {rapid Communications}\BibitemShut {NoStop}%
\bibitem [{\citenamefont {Yang}\ \emph {et~al.}(2025)\citenamefont {Yang},
  \citenamefont {Jiang}, \citenamefont {Wei}, \citenamefont {Wang},\ and\
  \citenamefont {Pan}}]{Jiang_3D}%
  \BibitemOpen
  \bibfield  {author} {\bibinfo {author} {\bibfnamefont {X.}~\bibnamefont
  {Yang}}, \bibinfo {author} {\bibfnamefont {X.-P.}\ \bibnamefont {Jiang}},
  \bibinfo {author} {\bibfnamefont {Z.}~\bibnamefont {Wei}}, \bibinfo {author}
  {\bibfnamefont {Y.}~\bibnamefont {Wang}},\ and\ \bibinfo {author}
  {\bibfnamefont {L.}~\bibnamefont {Pan}},\ }\bibfield  {title} {\bibinfo
  {title} {Dissipation-induced transition between delocalization and
  localization in the three-dimensional anderson model},\ }\href
  {https://doi.org/10.1103/PhysRevB.111.134203} {\bibfield  {journal} {\bibinfo
   {journal} {Phys. Rev. B}\ }\textbf {\bibinfo {volume} {111}},\ \bibinfo
  {pages} {134203} (\bibinfo {year} {2025})}\BibitemShut {NoStop}%
\bibitem [{\citenamefont {Liu}\ \emph {et~al.}(2024{\natexlab{c}})\citenamefont
  {Liu}, \citenamefont {Wang}, \citenamefont {Yang}, \citenamefont {Jie},\ and\
  \citenamefont {Wang}}]{WYC_PRL}%
  \BibitemOpen
  \bibfield  {author} {\bibinfo {author} {\bibfnamefont {Y.}~\bibnamefont
  {Liu}}, \bibinfo {author} {\bibfnamefont {Z.}~\bibnamefont {Wang}}, \bibinfo
  {author} {\bibfnamefont {C.}~\bibnamefont {Yang}}, \bibinfo {author}
  {\bibfnamefont {J.}~\bibnamefont {Jie}},\ and\ \bibinfo {author}
  {\bibfnamefont {Y.}~\bibnamefont {Wang}},\ }\bibfield  {title} {\bibinfo
  {title} {Dissipation-induced extended-localized transition},\ }\href
  {https://doi.org/10.1103/PhysRevLett.132.216301} {\bibfield  {journal}
  {\bibinfo  {journal} {Phys. Rev. Lett.}\ }\textbf {\bibinfo {volume} {132}},\
  \bibinfo {pages} {216301} (\bibinfo {year} {2024}{\natexlab{c}})}\BibitemShut
  {NoStop}%
\bibitem [{\citenamefont {Xu}\ \emph {et~al.}(2026)\citenamefont {Xu},
  \citenamefont {Wei}, \citenamefont {Jiang},\ and\ \citenamefont
  {Pan}}]{Xu_FlatBand}%
  \BibitemOpen
  \bibfield  {author} {\bibinfo {author} {\bibfnamefont {M.}~\bibnamefont
  {Xu}}, \bibinfo {author} {\bibfnamefont {Z.}~\bibnamefont {Wei}}, \bibinfo
  {author} {\bibfnamefont {X.-P.}\ \bibnamefont {Jiang}},\ and\ \bibinfo
  {author} {\bibfnamefont {L.}~\bibnamefont {Pan}},\ }\bibfield  {title}
  {\bibinfo {title} {Dissipation induced localization-delocalization transition
  in flat band systems},\ }\href@noop {} {\bibfield  {journal} {\bibinfo
  {journal} {iScience}\ }\textbf {\bibinfo {volume} {29}} (\bibinfo {year}
  {2026})}\BibitemShut {NoStop}%
\bibitem [{\citenamefont {Feng}\ \emph {et~al.}(2025)\citenamefont {Feng},
  \citenamefont {Zhou}, \citenamefont {Lu}, \citenamefont {Xianlong},\ and\
  \citenamefont {Cheng}}]{feng2025localization}%
  \BibitemOpen
  \bibfield  {author} {\bibinfo {author} {\bibfnamefont {X.}~\bibnamefont
  {Feng}}, \bibinfo {author} {\bibfnamefont {A.}~\bibnamefont {Zhou}}, \bibinfo
  {author} {\bibfnamefont {F.}~\bibnamefont {Lu}}, \bibinfo {author}
  {\bibfnamefont {G.}~\bibnamefont {Xianlong}},\ and\ \bibinfo {author}
  {\bibfnamefont {S.}~\bibnamefont {Cheng}},\ }\bibfield  {title} {\bibinfo
  {title} {Localization and topological properties of the nonequilibrium steady
  state in one-dimensional homogenous systems with disorder and dissipation},\
  }\href {https://doi.org/doi.org/10.1103/qn95-d4qy} {\bibfield  {journal}
  {\bibinfo  {journal} {Phys. Rev. B}\ }\textbf {\bibinfo {volume} {112}},\
  \bibinfo {pages} {104204} (\bibinfo {year} {2025})}\BibitemShut {NoStop}%
\bibitem [{\citenamefont {Roy}\ and\ \citenamefont
  {Gong}(2025)}]{roy2025aperiodic}%
  \BibitemOpen
  \bibfield  {author} {\bibinfo {author} {\bibfnamefont {S.}~\bibnamefont
  {Roy}}\ and\ \bibinfo {author} {\bibfnamefont {J.}~\bibnamefont {Gong}},\
  }\bibfield  {title} {\bibinfo {title} {Aperiodic dissipation as a mechanism
  for steady-state localization},\ }\href
  {https://doi.org/doi.org/10.1103/qprv-gf5g} {\bibfield  {journal} {\bibinfo
  {journal} {Phys. Rev. B}\ }\textbf {\bibinfo {volume} {112}},\ \bibinfo
  {pages} {155409} (\bibinfo {year} {2025})}\BibitemShut {NoStop}%
\bibitem [{\citenamefont {Hu}\ \emph {et~al.}(2025)\citenamefont {Hu},
  \citenamefont {Yang},\ and\ \citenamefont {Wang}}]{WYC_MBL}%
  \BibitemOpen
  \bibfield  {author} {\bibinfo {author} {\bibfnamefont {Y.}~\bibnamefont
  {Hu}}, \bibinfo {author} {\bibfnamefont {C.}~\bibnamefont {Yang}},\ and\
  \bibinfo {author} {\bibfnamefont {Y.}~\bibnamefont {Wang}},\ }\bibfield
  {title} {\bibinfo {title} {Inducing a transition between thermal and
  many-body localized states and detecting many-body mobility edges through
  dissipation},\ }\href {https://doi.org/10.1103/lyr8-n4r9} {\bibfield
  {journal} {\bibinfo  {journal} {Physical Review B}\ }\textbf {\bibinfo
  {volume} {111}},\ \bibinfo {pages} {174204} (\bibinfo {year}
  {2025})}\BibitemShut {NoStop}%
\bibitem [{\citenamefont {Wang}\ \emph
  {et~al.}(2024{\natexlab{c}})\citenamefont {Wang}, \citenamefont {Yuan},
  \citenamefont {Zhang}, \citenamefont {Wang}, \citenamefont {Deng},\ and\
  \citenamefont {Duan}}]{Diss_Scar1}%
  \BibitemOpen
  \bibfield  {author} {\bibinfo {author} {\bibfnamefont {H.-R.}\ \bibnamefont
  {Wang}}, \bibinfo {author} {\bibfnamefont {D.}~\bibnamefont {Yuan}}, \bibinfo
  {author} {\bibfnamefont {S.-Y.}\ \bibnamefont {Zhang}}, \bibinfo {author}
  {\bibfnamefont {Z.}~\bibnamefont {Wang}}, \bibinfo {author} {\bibfnamefont
  {D.-L.}\ \bibnamefont {Deng}},\ and\ \bibinfo {author} {\bibfnamefont
  {L.-M.}\ \bibnamefont {Duan}},\ }\bibfield  {title} {\bibinfo {title}
  {Embedding quantum many-body scars into decoherence-free subspaces},\ }\href
  {https://doi.org/10.1103/PhysRevLett.132.150401} {\bibfield  {journal}
  {\bibinfo  {journal} {Physical Review Letters}\ }\textbf {\bibinfo {volume}
  {132}},\ \bibinfo {pages} {150401} (\bibinfo {year}
  {2024}{\natexlab{c}})}\BibitemShut {NoStop}%
\bibitem [{\citenamefont {Shen}\ \emph {et~al.}(2024)\citenamefont {Shen},
  \citenamefont {Qin}, \citenamefont {Desaules}, \citenamefont {Papi{\'c}},\
  and\ \citenamefont {Lee}}]{Diss_Scar2}%
  \BibitemOpen
  \bibfield  {author} {\bibinfo {author} {\bibfnamefont {R.}~\bibnamefont
  {Shen}}, \bibinfo {author} {\bibfnamefont {F.}~\bibnamefont {Qin}}, \bibinfo
  {author} {\bibfnamefont {J.-Y.}\ \bibnamefont {Desaules}}, \bibinfo {author}
  {\bibfnamefont {Z.}~\bibnamefont {Papi{\'c}}},\ and\ \bibinfo {author}
  {\bibfnamefont {C.~H.}\ \bibnamefont {Lee}},\ }\bibfield  {title} {\bibinfo
  {title} {Enhanced many-body quantum scars from the non-hermitian fock skin
  effect},\ }\href {https://doi.org/10.1103/PhysRevLett.132.150401} {\bibfield
  {journal} {\bibinfo  {journal} {Phys. Rev. Lett.}\ }\textbf {\bibinfo
  {volume} {133}},\ \bibinfo {pages} {216601} (\bibinfo {year}
  {2024})}\BibitemShut {NoStop}%
\bibitem [{\citenamefont {Jiang}\ \emph {et~al.}(2025)\citenamefont {Jiang},
  \citenamefont {Xu}, \citenamefont {Yang}, \citenamefont {Hou}, \citenamefont
  {Wang},\ and\ \citenamefont {Pan}}]{Diss_scar3}%
  \BibitemOpen
  \bibfield  {author} {\bibinfo {author} {\bibfnamefont {X.-P.}\ \bibnamefont
  {Jiang}}, \bibinfo {author} {\bibfnamefont {M.}~\bibnamefont {Xu}}, \bibinfo
  {author} {\bibfnamefont {X.}~\bibnamefont {Yang}}, \bibinfo {author}
  {\bibfnamefont {H.}~\bibnamefont {Hou}}, \bibinfo {author} {\bibfnamefont
  {Y.}~\bibnamefont {Wang}},\ and\ \bibinfo {author} {\bibfnamefont
  {L.}~\bibnamefont {Pan}},\ }\bibfield  {title} {\bibinfo {title} {Robustness
  of quantum many-body scars in the presence of markovian bath},\ }\href@noop
  {} {\bibfield  {journal} {\bibinfo  {journal} {arXiv preprint}\ } (\bibinfo
  {year} {2025})},\ \Eprint {https://arxiv.org/abs/2501.00886}
  {arXiv:2501.00886} \BibitemShut {NoStop}%
\bibitem [{\citenamefont {Ma}\ \emph {et~al.}(2025)\citenamefont {Ma},
  \citenamefont {Guo}, \citenamefont {Gao}, \citenamefont {Papi{\'c}},\ and\
  \citenamefont {Ying}}]{Diss_scar4}%
  \BibitemOpen
  \bibfield  {author} {\bibinfo {author} {\bibfnamefont {J.-L.}\ \bibnamefont
  {Ma}}, \bibinfo {author} {\bibfnamefont {Z.}~\bibnamefont {Guo}}, \bibinfo
  {author} {\bibfnamefont {Y.}~\bibnamefont {Gao}}, \bibinfo {author}
  {\bibfnamefont {Z.}~\bibnamefont {Papi{\'c}}},\ and\ \bibinfo {author}
  {\bibfnamefont {L.}~\bibnamefont {Ying}},\ }\bibfield  {title} {\bibinfo
  {title} {Liouvillian spectral transition in noisy quantum many-body scars},\
  }\href {https://doi.org/doi.org/10.1103/4my3-vk6c} {\bibfield  {journal}
  {\bibinfo  {journal} {Phys. Rev. Lett.}\ }\textbf {\bibinfo {volume} {135}},\
  \bibinfo {pages} {180401} (\bibinfo {year} {2025})}\BibitemShut {NoStop}%
\bibitem [{\citenamefont {Garc{\'\i}a-Garc{\'\i}a}\ \emph
  {et~al.}(2025)\citenamefont {Garc{\'\i}a-Garc{\'\i}a}, \citenamefont {Lu},
  \citenamefont {S{\'a}},\ and\ \citenamefont {Verbaarschot}}]{Diss_scar5}%
  \BibitemOpen
  \bibfield  {author} {\bibinfo {author} {\bibfnamefont {A.~M.}\ \bibnamefont
  {Garc{\'\i}a-Garc{\'\i}a}}, \bibinfo {author} {\bibfnamefont
  {Z.}~\bibnamefont {Lu}}, \bibinfo {author} {\bibfnamefont {L.}~\bibnamefont
  {S{\'a}}},\ and\ \bibinfo {author} {\bibfnamefont {J.~J.}\ \bibnamefont
  {Verbaarschot}},\ }\bibfield  {title} {\bibinfo {title} {Lindblad many-body
  scars},\ }\href@noop {} {\bibfield  {journal} {\bibinfo  {journal} {arXiv
  preprint}\ } (\bibinfo {year} {2025})},\ \Eprint
  {https://arxiv.org/abs/2503.06665} {arXiv:2503.06665} \BibitemShut {NoStop}%
\bibitem [{\citenamefont {Dong}\ \emph {et~al.}(2025)\citenamefont {Dong},
  \citenamefont {Mu}, \citenamefont {Qin},\ and\ \citenamefont {Cui}}]{QME6}%
  \BibitemOpen
  \bibfield  {author} {\bibinfo {author} {\bibfnamefont {J.~W.}\ \bibnamefont
  {Dong}}, \bibinfo {author} {\bibfnamefont {H.~F.}\ \bibnamefont {Mu}},
  \bibinfo {author} {\bibfnamefont {M.}~\bibnamefont {Qin}},\ and\ \bibinfo
  {author} {\bibfnamefont {H.~T.}\ \bibnamefont {Cui}},\ }\bibfield  {title}
  {\bibinfo {title} {Quantum mpemba effect of localization in the dissipative
  mosaic model},\ }\href {https://doi.org/10.1103/PhysRevA.111.022215}
  {\bibfield  {journal} {\bibinfo  {journal} {Phys. Rev. A}\ }\textbf {\bibinfo
  {volume} {111}},\ \bibinfo {pages} {022215} (\bibinfo {year}
  {2025})}\BibitemShut {NoStop}%
\bibitem [{\citenamefont {Xu}\ \emph {et~al.}(2025{\natexlab{b}})\citenamefont
  {Xu}, \citenamefont {Wei}, \citenamefont {Jiang},\ and\ \citenamefont
  {Pan}}]{QME7}%
  \BibitemOpen
  \bibfield  {author} {\bibinfo {author} {\bibfnamefont {M.}~\bibnamefont
  {Xu}}, \bibinfo {author} {\bibfnamefont {Z.}~\bibnamefont {Wei}}, \bibinfo
  {author} {\bibfnamefont {X.-P.}\ \bibnamefont {Jiang}},\ and\ \bibinfo
  {author} {\bibfnamefont {L.}~\bibnamefont {Pan}},\ }\bibfield  {title}
  {\bibinfo {title} {Expedited thermalization dynamics in incommensurate
  systems},\ }\href {https://doi.org/10.1103/9qkm-35y1} {\bibfield  {journal}
  {\bibinfo  {journal} {Phys. Rev. A}\ }\textbf {\bibinfo {volume} {112}},\
  \bibinfo {pages} {042210} (\bibinfo {year} {2025}{\natexlab{b}})}\BibitemShut
  {NoStop}%
\bibitem [{\citenamefont {Zhou}\ \emph {et~al.}(2025)\citenamefont {Zhou},
  \citenamefont {Lu}, \citenamefont {Cheng},\ and\ \citenamefont
  {Gao}}]{QME705}%
  \BibitemOpen
  \bibfield  {author} {\bibinfo {author} {\bibfnamefont {A.}~\bibnamefont
  {Zhou}}, \bibinfo {author} {\bibfnamefont {F.}~\bibnamefont {Lu}}, \bibinfo
  {author} {\bibfnamefont {S.}~\bibnamefont {Cheng}},\ and\ \bibinfo {author}
  {\bibfnamefont {X.}~\bibnamefont {Gao}},\ }\bibfield  {title} {\bibinfo
  {title} {Quantum otto heat engine and quantum mpemba effect in quasiperiodic
  systems},\ }\href@noop {} {\bibfield  {journal} {\bibinfo  {journal} {arXiv
  preprint}\ } (\bibinfo {year} {2025})},\ \Eprint
  {https://arxiv.org/abs/2509.12572} {arXiv:2509.12572} \BibitemShut {NoStop}%
\bibitem [{\citenamefont {Wang}\ \emph {et~al.}(2020)\citenamefont {Wang},
  \citenamefont {Xia}, \citenamefont {Zhang}, \citenamefont {Yao},
  \citenamefont {Chen}, \citenamefont {You}, \citenamefont {Zhou},\ and\
  \citenamefont {Liu}}]{Mosaic1}%
  \BibitemOpen
  \bibfield  {author} {\bibinfo {author} {\bibfnamefont {Y.}~\bibnamefont
  {Wang}}, \bibinfo {author} {\bibfnamefont {X.}~\bibnamefont {Xia}}, \bibinfo
  {author} {\bibfnamefont {L.}~\bibnamefont {Zhang}}, \bibinfo {author}
  {\bibfnamefont {H.}~\bibnamefont {Yao}}, \bibinfo {author} {\bibfnamefont
  {S.}~\bibnamefont {Chen}}, \bibinfo {author} {\bibfnamefont {J.}~\bibnamefont
  {You}}, \bibinfo {author} {\bibfnamefont {Q.}~\bibnamefont {Zhou}},\ and\
  \bibinfo {author} {\bibfnamefont {X.}~\bibnamefont {Liu}},\ }\bibfield
  {title} {\bibinfo {title} {One-dimensional quasiperiodic mosaic lattice with
  exact mobility edges},\ }\href
  {https://doi.org/10.1103/PhysRevLett.125.196604} {\bibfield  {journal}
  {\bibinfo  {journal} {Phys. Rev. Lett.}\ }\textbf {\bibinfo {volume} {125}},\
  \bibinfo {pages} {196604} (\bibinfo {year} {2020})}\BibitemShut {NoStop}%
\bibitem [{\citenamefont {Deng}\ \emph {et~al.}(2020)\citenamefont {Deng},
  \citenamefont {Masella}, \citenamefont {Pupillo},\ and\ \citenamefont
  {Santos}}]{LongRange1}%
  \BibitemOpen
  \bibfield  {author} {\bibinfo {author} {\bibfnamefont {X.}~\bibnamefont
  {Deng}}, \bibinfo {author} {\bibfnamefont {G.}~\bibnamefont {Masella}},
  \bibinfo {author} {\bibfnamefont {G.}~\bibnamefont {Pupillo}},\ and\ \bibinfo
  {author} {\bibfnamefont {L.}~\bibnamefont {Santos}},\ }\bibfield  {title}
  {\bibinfo {title} {Universal algebraic growth of entanglement entropy in
  many-body localized systems with power-law interactions},\ }\href
  {https://doi.org/10.1103/PhysRevLett.125.010401} {\bibfield  {journal}
  {\bibinfo  {journal} {Phys. Rev. Lett.}\ }\textbf {\bibinfo {volume} {125}},\
  \bibinfo {pages} {010401} (\bibinfo {year} {2020})}\BibitemShut {NoStop}%
\bibitem [{\citenamefont {Lindblad}(1976)}]{Lindblad1}%
  \BibitemOpen
  \bibfield  {author} {\bibinfo {author} {\bibfnamefont {G.}~\bibnamefont
  {Lindblad}},\ }\bibfield  {title} {\bibinfo {title} {On the generators of
  quantum dynamical semigroups},\ }\href {https://doi.org/10.1007/BF01608499}
  {\bibfield  {journal} {\bibinfo  {journal} {Commun. Math. Phys.}\ }\textbf
  {\bibinfo {volume} {119}},\ \bibinfo {pages} {48} (\bibinfo {year}
  {1976})}\BibitemShut {NoStop}%
\bibitem [{\citenamefont {Gorini}\ \emph {et~al.}(1976)\citenamefont {Gorini},
  \citenamefont {Kossakowski},\ and\ \citenamefont {Sudarsahan}}]{Lindblad2}%
  \BibitemOpen
  \bibfield  {author} {\bibinfo {author} {\bibfnamefont {V.}~\bibnamefont
  {Gorini}}, \bibinfo {author} {\bibfnamefont {A.}~\bibnamefont
  {Kossakowski}},\ and\ \bibinfo {author} {\bibfnamefont {E.~C.}\ \bibnamefont
  {Sudarsahan}},\ }\bibfield  {title} {\bibinfo {title} {Completely positive
  dynamical semigroups of n-level systems},\ }\href
  {https://doi.org/10.1063/1.522979} {\bibfield  {journal} {\bibinfo  {journal}
  {J. Math. Phys.}\ }\textbf {\bibinfo {volume} {17}},\ \bibinfo {pages} {821}
  (\bibinfo {year} {1976})}\BibitemShut {NoStop}%
\bibitem [{\citenamefont {Moy}\ \emph {et~al.}(1999)\citenamefont {Moy},
  \citenamefont {Hope},\ and\ \citenamefont {Savage}}]{Moy1999}%
  \BibitemOpen
  \bibfield  {author} {\bibinfo {author} {\bibfnamefont {G.~M.}\ \bibnamefont
  {Moy}}, \bibinfo {author} {\bibfnamefont {J.~J.}\ \bibnamefont {Hope}},\ and\
  \bibinfo {author} {\bibfnamefont {C.~M.}\ \bibnamefont {Savage}},\ }\bibfield
   {title} {\bibinfo {title} {Born and markov approximations for atom lasers},\
  }\href {https://doi.org/10.1103/PhysRevA.59.667} {\bibfield  {journal}
  {\bibinfo  {journal} {Phys. Rev. A}\ }\textbf {\bibinfo {volume} {59}},\
  \bibinfo {pages} {667} (\bibinfo {year} {1999})}\BibitemShut {NoStop}%
\bibitem [{\citenamefont {Breuer}\ and\ \citenamefont
  {Petruccione}(2002)}]{Breuer2002}%
  \BibitemOpen
  \bibfield  {author} {\bibinfo {author} {\bibfnamefont {H.-P.}\ \bibnamefont
  {Breuer}}\ and\ \bibinfo {author} {\bibfnamefont {F.}~\bibnamefont
  {Petruccione}},\ }\href@noop {} {\emph {\bibinfo {title} {The Theory of Open
  Quantum Systems}}}\ (\bibinfo  {publisher} {Oxford University Press},\
  \bibinfo {address} {Oxford},\ \bibinfo {year} {2002})\BibitemShut {NoStop}%
\end{thebibliography}%
\end{document}